\documentclass[prd,preprintnumbers,amsmath,amssymb,floatfix]{revtex4}
\usepackage{graphicx}
\usepackage{bm}
\usepackage{amsfonts}
\usepackage{lineno,hyperref}
\usepackage{array}
\usepackage{color,multirow,cases, makecell}
\usepackage{pifont, float, setspace, moresize}

\begin{document}

\title{Polarization modes of gravitational waves in general Einstein-vector theory}

\author{Xiao-Bin Lai$^{a}$$^{b}$}
\author{Yu-Qi Dong$^{a}$$^{b}$}
\author{Yu-Qiang Liu$^{a}$$^{b}$}
\author{Yu-Xiao Liu$^{a}$$^{b}$\footnote{liuyx@lzu.edu.cn}}

\affiliation{
$^{a}$Lanzhou Center for Theoretical Physics,
  Key Laboratory of Quantum Theory and Applications of the Ministry of Education,
  Key Laboratory of Theoretical Physics of Gansu Province,
  School of Physical Science and Technology,
  Lanzhou University, Lanzhou 730000, China \\
$^{b}$Institute of Theoretical Physics $\&$ Research Center of Gravitation,
  Lanzhou University, Lanzhou 730000, China
}

\begin{abstract}
We study the polarization modes of gravitational waves in general Einstein-vector theory with an arbitrary constant background vector field under a Minkowski background. We compare these polarization modes with those of other vector-tensor theories and constrain the parameter spaces based on the gravitational-wave event GW170817 with its electromagnetic counterpart GRB170817A and observations of pulsar timing arrays.
The presence of the background vector field leads to the anisotropy of space and a rich variety of gravitational wave contents.
Our results reveal that the polarization modes of gravitational waves depend on the parameter spaces.
There are at least two and at most five independent polarization modes in one parameter space. In different parameter spaces, some mixture modes are allowed, including tensor-vector, tensor-scalar, tensor-vector-scalar, vector-scalar, and scalar-scalar mixture modes, as well as five independent modes (excluding $P_l$): $P_+$, $P_{\times}$, $P_x$, $P_y$, and $P_b$. 
In all regions of the parameter spaces, there are always two tensor modes, which can be either independent of or mixed with other modes. The independent $P_b$ mode consistently exhibits the same speed as light.
If the speed of tensorial gravitational waves strictly equals that of light, only the $P_+$, $P_{\times}$, and $P_b$ modes are permitted. 
Furthermore, through comparisons of some vector-tensor theories and based on the observations of gravitational waves, the $P_b$ mode is expected to be allowed.
\end{abstract}

\maketitle
\tableofcontents 

\section{Introduction}
Since the first direct detection of gravitational waves (GWs) by Advanced Laser Interferometer Gravitational-wave Observatory (LIGO) 
in 2015~\cite{LIGOScientific:2016aoc,LIGOScientific:2016sjg}, the bridge between theory and experiment of GWs
has been built. It implies that we have entered the era of gravitational-wave astronomy~\cite{Zhao:2021zlr, Bi:2023tib, Wang:2023len, Wang:2021srv, Gao:2022hho, Shan:2022xfx}. Subsequently, with an increasing number of gravitational-wave signals detected by Advanced LIGO and Virgo~\cite{LIGOScientific:2017bnn,LIGOScientific:2017ycc,LIGOScientific:2017vwq,LIGOScientific:2021qlt,LIGOScientific:2018mvr,LIGOScientific:2020ibl,LIGOScientific:2021djp}, 
abundant detection resources have been provided for theoretical research of testing gravity theories and understanding the Universe. 
Meanwhile, among the fundamental characteristics of GWs, their polarization modes have emerged as strong candidates for the study of GWs and for testing theories of gravity.
Last year, the data, collected by the North American Nanoherz Observatory for GWs, the European Pulsar Timing Array, the Indian Pulsar Timing Array, the Parkes Pulsar Timing Array, and the Chinese 
Pulsar Timing Array colleagues, revealed evidence of a stochastic gravitational-wave background at nano-hertz frequencies~\cite{NANOGrav:2023gor,EPTA:2023fyk,Reardon:2023gzh,Xu:2023wog}. It implies pulsar timing array (PTA) may open new windows to explore the nano-hertz 
frequency range of the spectrum, and it will give us another way to detect the polarization modes of GWs.\par

Einstein's general relativity, the most widely accepted theory of gravity, has undergone extensive testing in various ways, particularly within the Solar 
System. However, there is suspicion within the physics community that it may not be the ultimate theory of gravity. This is due to the problems 
related to dark matter and dark energy~\cite{Smith:1936mlg,zwicky1933helvetica,Zwicky:1937zza,Peebles:2002gy}, quantization~\cite{Goroff:1985th,tHooft:1974toh}, and other challenges~\cite{SupernovaCosmologyProject:1998vns,La:1989za}. In the pursuit of a more 
fundamental theory of gravity, many modified gravity theories have been 
proposed. 
One significant approach to modified gravity in four dimensions involves the introduction of additional fields that minimally or nonminimally couple to curvature. 
Scalar-tensor theory is one such category in which an additional scalar field nonminimally couples to curvature, as seen in Brans-Dicke theory~\cite{PhysRev.124.925}, 
Horndeski theory~\cite{Horndeski:1974wa,Fu:2019xtx}, and others~\cite{DeFelice:2010jn}. 
Similarly, vector-tensor theory encompasses theories such as Einstein-æther 
theory~\cite{Jacobson:2007veq}, generalized Proca theory~\cite{Tasinato:2014eka, Heisenberg:2014rta}, Bumblebee theory~\cite{Kostelecky:2003fs}, and general Einstein-vector theory~\cite{Geng:2015kvs}. Furthermore, there are theories like tensor-vector-scalar theory (TeVeS)~\cite{Bekenstein:2004ne} and scalar-tensor-vector 
theory~\cite{Moffat:2005si}. 
Another popular method for modifying general relativity is to treat the metric and the connection as independent dynamic variables, known as the Palatini formalism. 
These theories are of particular interest due to their predictive capabilities in cosmic inflation~\cite{Bauer:2008zj,Markkanen:2017tun} and for other aspects~\cite{Vollick:2003aw,Bauer:2010jg,Rubio:2019ypq}. 
There exist many modified theories within the Palatini formalism as well~\cite{Olmo:2011uz, BeltranJimenez:2018vdo,Dong:2021jtd,Lu:2020eux,Kubota:2020ehu}. 
In addition to the above two approaches, there are many other popular gravity theories, including the higher derivative theories like $f(R)$ theory~\cite{Sotiriou:2008rp,Cui:2020fiz,Baibosunov:1990qm,Chen:2020zzs}, higher-dimensional theories such as Kaluza-Klein theory~\cite{Kaluza:1921tu}, and others~\cite{Rubakov:1983bb,Tan:2020sys,Arkani-Hamed:1998jmv,Randall:1999vf,Yu:2019jlb,Liu:2017gcn,Li:2022kly,Lin:2022hus,Xu:2022xxd,Zhong:2022wlw}. All these theories 
have made significant contributions to our understanding of gravity.\par

Polarization is one of the most crucial aspects of GWs. It can be the most direct evidence to 
test gravity theories. General relativity predicts there are two tensor polarization modes at the same speed as light: plus mode 
($P_+$) and cross mode ($P_{\times}$). The tensor modes of GWs have been confirmed at first by the detection of GW170814~\cite{LIGOScientific:2017ycc}. The speed of GWs has been bounded tightly by detections and analysis~\cite{LIGOScientific:2017vwq, Goldstein:2017mmi,Savchenko:2017ffs}. However, in the most general metric theory, there are all six independent modes in four-dimensional space-time~\cite{Eardley:1973zuo}: two transverse tensor modes (plus $P_+$ and cross $P_{\times}$), two vector modes (vector-$x$ $P_{x}$ and vector-$y$ $P_y$), one transverse scalar mode (breathing $P_{b}$), and one longitudinal scalar mode (longitudinal $P_l$). In a specific metric gravity theory, a subset of these modes or various mixtures of them may exist.
Usually, the plus and cross modes should be included in order to ensure consistency with general relativity under certain limits.
The study of the polarizations in modified gravity theories has always been a significant topic in the gravity community. 
There are various approaches to analyze modified theories for their polarizations, such as the Newman-Penrose (NP) formalism~\cite{Eardley:1973zuo,Eardley:1973br,Hyun:2018pgn} and the Bardeen framework~\cite{Bardeen:1980kt,Flanagan:2005yc}. Now, the 
polarization modes of many modified theories have been obtained~\cite{Liang:2022hxd, Dong:2021jtd, Dong:2023xyb, Liang:2017ahj, Gong:2018vbo, Wagle:2019mdq, Bahamonde:2021dqn, Lu:2020eux},
and some studies about GWs are under way~\cite{Dong:2023bgt, Liu:2022cwb, Dong:2022cvf, Kuroda:2015owv, Isi:2017equ, Niu:2018oox, Hou:2024xbv, Battista:2021rlh, Gao:2021vxb, Wu:2023hsa, Yi:2023mbm, Bian:2021ini, Schumacher:2023cxh}. Recently, the data of PTAs indicates a 
high probability that a breathing mode has been detected~\cite{Chen:2023uiz}. This deepens the interest of the 
gravity community in modified gravity theories and the polarizations of GWs once again.\par

In this paper, we will analyze the polarizations of GWs in general Einstein-vector theory with an arbitrary constant background vector field under a Minkowski background. 
We will derive the dynamical equations for gravity and the vector field. Subsequently, we will consider the perturbations of the dynamical 
variables (metric $g_{\mu\nu}$ and vector $A^{\mu}$) under the Minkowski background, and obtain the background field equations and perturbation equations.
These background field equations will enable us to constrain the parameters and the background vector field in this theory. For the convenience of calculation and analysis, 
we will set the propagation direction of GWs as ``$+z$'' and choose coordinate such that the $y$-component of the background vector field is zero. 
Then we will analyze the polarizations of GWs and their amplitude relationships based on the parameter spaces and the background vector field.
Since the vector field leads to the loss of space-time symmetry, 
the space is anisotropic, and the rotation symmetry is broken.
We will apply the Bardeen framework~\cite{Bardeen:1980kt,Flanagan:2005yc} to analyze the polarizations of GWs in general Einstein-vector theory. Afterward, we will briefly introduce the polarization modes of Einstein-æther theory, generalized Proca theory, and Bumblebee gravity theory, and compare them with general Einstein-vector theory. Finally, we will analyze the polarization modes and restrict the parameter spaces of general Einstein-vector theory based on the gravitational-wave event GW170817 with its electromagnetic counterpart GRB170817A and observations of PTAs.\par

This paper is structured as follows. In Sec.~\ref{GEVT}, we provide a briefly introduction to general Einstein-vector theory and present the field equations. Then we obtain the background field equations and the linear perturbation equations.
In Sec.~\ref{section-Gp}, we provide a concise overview of the Bardeen framework and present the relationships between gauge invariants and polarization modes.
In Sec.~\ref{P-modes}, we analyze the polarization modes and their amplitude relations of GWs in general Einstein-vector theory with an arbitrary constant background vector field.
In Sec.~\ref{analysis}, we summarize the results of Sec.~\ref{P-modes} and compare general Einstein-vector theory with Einstein-æther theory, generalized Proca theory, and Bumblebee gravity theory. Then we analyze the polarization modes and restrict the parameter spaces based on the observations of GWs.
Finally, in Sec.~\ref{conclusion}, we present our conclusion. \par

Throughout this paper, we will only consider four-dimensional space-time.
We use the following conventions: Greek letters $(\mu,\nu,\alpha,\beta,\cdots)$ in index list the spacetime coordinates, Latin letters $(i,j,k,\cdots)$ in index only list the spatial coordinates, the metric signature is $(-,+,+,+)$, and the speed of light is set as c = 1.

\section{General Einstein-vector theory and linear perturbation equations} \label{GEVT}
General Einstein-vector theory is a vector-tensor theory in general $D$ dimensions~\cite{Geng:2015kvs}. 
In addition to the metric $g_{\mu\nu}$, the theory involves one vector field $A^{\mu}$ which couples bilinearly to curvature. 
And the curvature terms are at most $\frac{D}{2}$ (or $\frac{D-1}{2}$ when $D$ is odd) order of curvature combined by some way~\cite{Geng:2015kvs,lovelock1971einstein}. 
In this theory, the vector field is linear in its equation of motion, and for curvature, it is Riemann tensor rather than its 
derivative that enters the equations of motion. We will briefly introduce this theory and obtain its linear perturbation equations in this section. 

\subsection{General Einstein-vector theory}
The Lagrangian of general Einstein-vector theory is~\cite{Geng:2015kvs} 
\begin{equation}
    \mathcal{L}=\sqrt{-g}\Big( -\frac{1}{4}F^2+\sum_{k=0}\big( \alpha^{(k)}E^{(k)}+\beta^{(k)}\widetilde{G}^{(k)}
    +\gamma^{(k)}G^{(k)} \big) \Big),\label{lagrangian_GEV}
\end{equation}
where $F^2=F_{\mu\nu}F^{\mu\nu}$, $F_{\mu\nu}=\nabla_{\mu}A_{\nu}-\nabla_{\nu}A_{\mu}$ is the field strength of 
the vector potential $A^{\mu}$,  $\alpha^{(k)}$, $\beta^{(k)}$, and $\gamma^{(k)}$ are 
series of constants, and $E^{(k)}$, $\widetilde{G}^{(k)}$, and $G^{(k)}$ are defined by
\begin{eqnarray}
    && E^{(k)}=\frac{1}{2^k}\delta^{\beta_1 \cdots \beta_{2k}}_{\alpha_1 \cdots \alpha_{2k}}
    R^{\alpha_1\alpha_2}_{\quad\;\;\;\beta_1\beta_2} \cdots R^{\alpha_{2k-1}\alpha_{2k}}_{\qquad\quad\;\;\beta_{2k-1}\beta_{2k}},\\
    && \widetilde{G}^{(k)}=E^{(k)}A^2,\\
    && G^{(k)}=E^{(k)}_{\mu\nu}A^{\mu}A^{\nu}.
\end{eqnarray}
Here, $E^{(k)\nu}_{\mu}=-\frac{1}{2^{k+1}}\delta^{\beta_1 \cdots \beta_{2k}\nu}_{\alpha_1 \cdots \alpha_{2k}\mu}
R^{\alpha_1\alpha_2}_{\quad\;\;\;\beta_1\beta_2} \cdots R^{\alpha_{2k-1}\alpha_{2k}}_{\qquad\quad\;\;\beta_{2k-1}\beta_{2k}}$, $R^{\mu\nu}_{\quad\alpha\beta}$ is Riemann tensor, $\delta^{\beta_1 \cdots \beta_{s}}_{\alpha_1 \cdots \alpha_{s}}=s!\delta^{\beta_1}_{[\alpha_1} \cdots \delta^{\beta_s}_{\alpha_s]}$, and 
$A^2=A_{\mu}A^{\mu}$.\par 
There are many similarities and differences between this theory and generalized Proca theory~\cite{Heisenberg:2014rta}. In linear approximations, curvature is linear in equations of motion, and the equations of motion are second order for both theories. 
The differences between these two theories are listed as follows:
\begin{itemize}
    \item The self-interaction of the vector field is considered in generalized Proca theory, instead of this theory.
    \item The derivative of the vector field couples to curvature only in generalized Proca theory. 
    \item Curvature is linear in the action of generalized Proca theory, but it depends on the dimension of space-time for this theory.
    \item The vector field is linear in its equations of motion in general Einstein-vector theory, but it is not in generalized Proca theory.
\end{itemize}

\indent In this paper, we focus on the polarizations of GWs in a four-dimensional Minkowski background. 
Since the terms with $k\ge 2$ have no contribution to the linear perturbation equations, we only need to consider $k=0$ and $k=1$. 
After resetting the parameters, the action is given by~\cite{Geng:2015kvs} 
\begin{eqnarray}
    S=\frac{1}{16\pi G}\int dx^{4}\sqrt{-g}\Big(R-2\Lambda_{0}-\frac{1}{4}F^{2}-\frac{1}{2}\mu_{0}^{2}A^{2}+\beta R A^{2}+\gamma G_{\mu\nu}A^{\mu}A^{\nu}\Big). \label{action2-2104}
\end{eqnarray}
Here, $G_{\mu\nu}=R_{\mu\nu}-\frac{1}{2}g_{\mu\nu}R$ is the Einstein tensor, $\Lambda_0$ is the cosmological constant, $\mu_0$ is the mass of the vector field, and
$\beta$ and $\gamma$ are two coupling constants of gravity and the vector field.\par
By varying the action with respect to the vector $A^{\mu}$ and the metric $g^{\mu\nu}$, we can obtain the equations of motion~\cite{Geng:2015kvs}:
\begin{eqnarray}
    \nabla_{\nu}F^{\nu\mu}&=&\mu_{0}^{2}A^{\mu}-2\beta R A^{\mu}-2\gamma A_{\nu}G^{\nu\mu},\label{motion-A}\\
    G_{\mu\nu}&=&-\Lambda_{0}g_{\mu\nu}+\frac{1}{2}\Big(F_{\mu\nu}^{2}-\frac{1}{4}g_{\mu\nu}F^{2}\Big)
    +\frac{1}{2}\mu_{0}^{2}\Big(A_{\mu}A_{\nu}-\frac{1}{2}g_{\mu\nu}A^{2}\Big)+\beta Y_{\mu\nu}+\gamma Z_{\mu\nu}, \label{motion-g}
\end{eqnarray}
where
\begin{eqnarray}
    Y_{\mu\nu}&=&-R A_{\mu}A_{\nu}-G_{\mu\nu}A^{2}+(\nabla_{\mu}\nabla_{\nu}-g_{\mu\nu}\square)A^{2}, \\
    Z_{\mu\nu}&=&\frac{1}{2}A^{2}R_{\mu\nu}+\frac{1}{2}R A_{\mu}A_{\nu}-2A^{\alpha}R_{\alpha(\mu}A_{\nu)}
    -\frac{1}{2}\nabla_{\mu}\nabla_{\nu}A^{2}-\frac{1}{2}\square (A_{\mu}A_{\nu})
    +\nabla_{\alpha}\nabla_{(\mu}(A_{\nu)}A^{\alpha}) \nonumber\\
    && +\frac{1}{2} g_{\mu\nu}\big(G_{\alpha\beta}A^{\alpha}A^{\beta}+\square A^{2}
    -\nabla_{\alpha}\nabla_{\beta}(A^{\alpha}A^{\beta})\big).
\end{eqnarray}
In Eq.~\eqref{motion-A}, the vector field is linear, and in both Eqs.~\eqref{motion-A} and~\eqref{motion-g}, curvature is also linear. 
It is easy to see that the two equations involve at most second-order derivatives. 

\subsection{Linear perturbation equations} \label{LPE}
In this paper, we study the polarization modes of GWs in a Minkowski background. So we perturb 
the dynamical variables under the Minkowski background:
\begin{eqnarray}
    g_{\mu\nu}=\eta_{\mu\nu}+h_{\mu\nu} ,\quad A^{\mu}=\mathring{A}^{\mu}+a^{\mu}.\label{perturbation-gA}
\end{eqnarray}
Here, $\eta_{\mu\nu}$ is the Minkowski metric, $\mathring{A}^{\mu}$ is the background vector field, and $\mathring{A}^{\mu}$ is a constant vector. In the following, we use the Minkowski metrics $\eta^{\mu\nu}$ and $\eta_{\mu\nu}$ 
to raise and lower the space-time indices of perturbed quantities.
For unperturbed quantities, we still use $g^{\mu\nu}$ and $g_{\mu\nu}$ to raise and lower the space-time indices while  maintaining them to linear order, 
for instance,
\begin{eqnarray}
    g^{\mu\nu}=\eta^{\mu\nu}-h^{\mu\nu},\quad
    A_{\mu}=\mathring{A}_{\mu}+a_{\mu}+\mathring{A}^{\nu} h_{\mu\nu}.
\end{eqnarray}
Here, the notion `` $\mathring{ }$ '' above the letters means the corresponding zeroth-order quantities of the perturbations. And $h_{\mu\nu}$, $h^{\mu\nu}$, $a_{\mu}$, and $a^{\mu}$ 
represent the corresponding first-order quantities of the perturbations. \par
By substituting the perturbations~\eqref{perturbation-gA} into the field equations~\eqref{motion-A} and~\eqref{motion-g} while preserving them to zero order, 
we can derive the background equations:
\begin{eqnarray}
    \mu_0^{2}\mathring{A}^{\mu}&=&0, \label{background1-1620}\\
    2\mu_0^{2}\mathring{A}_{\mu}\mathring{A}_{\nu}-\mu_0^{2}\eta_{\mu\nu}\mathring{A}^2-2\eta_{\mu\nu}\Lambda_0&=&0.\label{background2-1620}
\end{eqnarray}
In order to simultaneously fulfill the above two equations, the parameter spaces are limited to the following two cases:
\begin{eqnarray}
    &&\text{\ding{172}}\;\; \Lambda_0=0,\;\; \mathring{A}^{\mu}=0; \label{case1-1627}\\ 
    &&\text{\ding{173}}\;\; \Lambda_0=0,\;\; \mu_0=0. \label{case2-1627}
\end{eqnarray}

\textbf{Case \ding{172}: $\Lambda_0=0$, $\mathring{A}^{\mu}=0$}. Substituting the perturbations~\eqref{perturbation-gA} into the 
field equations~\eqref{motion-A} and~\eqref{motion-g}, we can obtain the perturbation equations:
\begin{eqnarray}
    \mu_{0}^{2}a^{\mu}+\partial_{\nu}\partial^{\mu}a^{\nu}-\overline{\square} a^{\mu}&=&0, \label{linear1-A}\\
    \overset{\text{{\fontsize{3.5pt}{\baselineskip}\selectfont \textbf{(1)}}}}{G}_{\mu\nu}&=&0, \label{linear1-g}
\end{eqnarray}
where $\overline{\square}=\partial_{\mu}\partial^{\mu}$ is d'Alembert operator, and $\overset{\text{{\fontsize{3.5pt}{\baselineskip}\selectfont \textbf{(1)}}}}{G}_{\mu\nu}$ represents the first-order perturbation term of $G_{\mu\nu}$. In this paper, the notion `` $^{\overset{\text{{\fontsize{3.5pt}{\baselineskip}\selectfont \textbf{(1)}}}}{}}$ '' above the letters represent the corresponding first-order quantities of the perturbations.
\par
\textbf{Case \ding{173}: $\Lambda_0=0$, $\mu_0=0$}. In this case, we can obtain the perturbation equations:
\begin{eqnarray}
    \partial^{\mu}\partial_{\nu}a^{\nu}-\overline{\square}a^{\mu}&=&2\gamma \mathring{A}_{\nu}\overset{\text{{\fontsize{3.5pt}{\baselineskip}\selectfont \textbf{(1)}\quad\quad\;\;}}}{G^{\nu\mu}}+2\beta\mathring{A}^{\mu}\overset{\text{{\fontsize{3.5pt}{\baselineskip}\selectfont \textbf{(1)}}}}{R}
    +\mathring{A}_{\alpha}\overline{\square}h^{\alpha\mu}-\mathring{A}_{\alpha}\partial^{\mu}\partial_{\nu}h^{\alpha\nu}, \label{perturbationEq-a}\\
    \overset{\text{{\fontsize{3.5pt}{\baselineskip}\selectfont \textbf{(1)}}}}{G}_{\mu\nu}&=&\beta \overset{\text{{\fontsize{3.5pt}{\baselineskip}\selectfont \textbf{(1)}}}}{Y}_{\mu\nu}+\gamma \overset{\text{{\fontsize{3.5pt}{\baselineskip}\selectfont \textbf{(1)}}}}{Z}_{\mu\nu}.\label{perturbationEq-g}
\end{eqnarray}
Here, $\overset{\text{{\fontsize{3.5pt}{\baselineskip}\selectfont \textbf{(1)}}}}{Y}_{\mu\nu}$ and $\overset{\text{{\fontsize{3.5pt}{\baselineskip}\selectfont \textbf{(1)}}}}{Z}_{\mu\nu}$ are the first-order perturbation quantities of $Y_{\mu\nu}$ and $Z_{\mu\nu}$, respectively.
 \par
In the following, we will analyze the polarizations of GWs based on these background and perturbation equations under Bardeen framework.

\section{Gauge invariants and polarization modes of GWs}\label{section-Gp}
Here, we will first provide a concise overview of the Bardeen framework~\cite{Bardeen:1980kt,Flanagan:2005yc}, which is a method for finding gauge invariants for metric theories under gauge transformations.
We will then present the relationships between polarization modes and gauge invariants. 
Since a linearized theory is invariant under gauge transformations, we can simplify the linear perturbation equations using gauge invariants. Through gauge transformations, we can 
eliminate spurious degrees of freedom and retain actual physical quantities: gauge invariants. These gauge invariants can describe all polarization modes of GWs.

\subsection{Gauge invariants}
We begin by introducing the decomposition of the metric perturbation and the vector perturbation $h_{\mu\nu}$ and $a^{\mu}$ into irreducible pieces~\cite{Flanagan:2005yc, Liang:2022hxd, jackiw2003chern}:
\begin{equation}
    \begin{cases}
        h_{tt}&=\;\; 2\phi_h, \\
        h_{ti}&=\;\; \lambda_i +\partial_i \rho, \\
        h_{ij}&=\;\; h^{TT}_{ij}+\frac{1}{3}\delta_{ij}H+\partial_{(i}\varepsilon_{j)}+\big( \partial_i\partial_j - \frac{1}{3}\delta_{ij}\overline{\nabla}^2\big)\alpha, \\
        a^t&=\;\;\phi_a, \\
        a^i&=\;\;\mu^i +\partial^i \omega,
    \end{cases} \label{decomposition-ha}
\end{equation}
together with the constraints
\begin{eqnarray}
        \partial_i\lambda^i=\partial_i\varepsilon^i=\partial_i\mu^i=0,\quad \partial^i h^{TT}_{ij}=\delta^{ij}h^{TT}_{ij}=0.
\end{eqnarray}
Here, $H=\delta^{ij}h_{ij}$, and $\overline{\nabla}^2=\delta^{ij}\partial_i\partial_j$. As we can see above, a four-dimensional vector $a^{\mu}$ can be divided into its temporal part and spatial part $a^{\mu}=(a^t,a^i)$.
The spatial part can be further decomposed into transverse and longitudinal spatial vectors. Similarly, a four-dimensional tensor can also be decomposed, and the spatial tensor $h_{ij}$ 
can be decomposed into three parts: transverse-transverse ($h^{TT}_{ij}$), longitudinal-transverse ($\partial_{(i}\varepsilon_{j)}$), and longitudinal-longitudinal 
$\big(\frac{1}{3}\delta_{ij}H+\big( \partial_i\partial_j - \frac{1}{3}\delta_{ij}\overline{\nabla}^2\big)\alpha\big)$.\par
Now, consider the gauge transformations $x_{\mu}\rightarrow x_{\mu}+\xi_{\mu}$. Under linear approximations, the transformations of the metric and the vector field are as follows~\cite{bluhm2008spontaneous}:
\begin{eqnarray}
    h_{\mu\nu}&\rightarrow& h_{\mu\nu}-\partial_{\mu}\xi_{\nu}-\partial_{\nu}\xi_{\mu},\label{transformation1-ha}\\
    a^{\mu}&\rightarrow& a^{\mu}+\mathring{A}^{\nu}\partial_{\nu}\xi^{\mu}. \label{transformation2-ha}
\end{eqnarray}
Here, $\xi_{\mu}$ are arbitrary functions of space-time coordinates, and their derivatives $\partial_{\mu}\xi_{\nu}$ are at most of the same order as $h_{\mu\nu}$. \par
We can also take the decomposition $\xi_{\mu}=(\xi_t,\xi_i)=(\xi_t,B_i+\partial_iC)$ with $\partial_iB_i=0$. And by substituting this decomposition and 
the transformations~\eqref{transformation1-ha} and~\eqref{transformation2-ha} into the decompositions~\eqref{decomposition-ha}, we can obtain a set of gauge invariants:
\begin{equation}
    \begin{cases}
        h^{TT}_{ij} &\equiv\;\; h^{TT}_{ij}, \\
        \Xi_i &\equiv\;\; \lambda_i-\frac{1}{2}\dot{\varepsilon}_i, \\
        \Phi &\equiv\;\; -\phi_h+\dot{\rho}-\frac{1}{2}\ddot{\alpha}, \\
        \Theta &\equiv\;\; \frac{1}{3}\big(H-\overline{\nabla}^2 \alpha\big), \\
        \Sigma_{i} &\equiv\;\; \mu_{i}+\frac{1}{2}\mathring{A}^{t}\dot{\varepsilon}_{i}+\frac{1}{2}\mathring{A}^{j}\partial_{j}\varepsilon_{i}, \\
        \Omega &\equiv\;\; \phi_{a}-\mathring{A}^{t}\dot{\rho}-\mathring{A}^{i}\partial_{i}\rho+\frac{1}{2}\mathring{A}^{t}\ddot{\alpha}+\frac{1}{2}\mathring{A}^{i}\partial_{i}\dot{\alpha}, \\
        \Psi &\equiv\;\; \omega+\frac{1}{2}\mathring{A}^{t}\dot{\alpha}+\frac{1}{2}\mathring{A}^{j}\partial_{j}\alpha,
    \end{cases}\label{invariants-all}
\end{equation}
where the notation ``$\cdot$'' above a letter denotes ``$\partial_t$''. It is not difficult to prove that both $\Xi_i$ and $\Sigma_i$ are spatial transverse vectors, 
$\partial_{i}\Xi^i=\partial_{i}\Sigma^i=0$. It can be seen that there are seven gauge invariants 
under linear approximations, including four scalars $(\Phi,\Theta,\Omega,\Psi)$, two vectors $(\Xi_i,\Sigma_i)$, and 
one tensor $(h^{TT}_{ij})$.

\subsection{Polarization modes of GWs}
Now, we  analyze the relationships between the gauge invariants~\eqref{invariants-all} and the polarization modes of GWs. 
According to the general principle of relativity, we can not identify GWs by detecting the motion of a single particle. Instead, we can detect GWs 
by observing the relative displacement of two adjacent free test particles. This ``relative displacement" is the specific manifestation of the polarizations of GWs, 
and is described by the geodesic deviation equation. 
Under the conditions of the weak field and the low speed, the geodesic deviation equation is
\begin{eqnarray}
    \frac{\mathrm{d}^2 \delta x^i}{\mathrm{d}t^2}=-\overset{\text{{\fontsize{3.5pt}{\baselineskip}\selectfont \textbf{(1)}\quad}}}{R^i}_{\;tjt}\delta x^j. \label{deviation-1455}
\end{eqnarray}
Here, $\delta x^i$ represents the relative displacement of the two adjacent test particles, and $\overset{\text{{\fontsize{3.5pt}{\baselineskip}\selectfont \textbf{(1)}\quad}}}{R^i}_{\;tjt}$ is the components of the Riemann tensor. The equation~\eqref{deviation-1455} illustrates that 
the relative behavior of free test particles, which serves as the signal of the polarizations of GWs, is completely determined by $\overset{\text{{\fontsize{3.5pt}{\baselineskip}\selectfont \textbf{(1)}\quad}}}{R^i}_{\;tjt}$. 
Under linear approximations, by applying the scalar-vector-tensor decomposition~\eqref{decomposition-ha} and the gauge invariants~\eqref{invariants-all}, we can obtain the 
specific form of $\overset{\text{{\fontsize{3.5pt}{\baselineskip}\selectfont \textbf{(1)}\quad}}}{R^i}_{\;tjt}$ consisting of the gauge invariants:
\begin{eqnarray}
    \overset{\text{{\fontsize{3.5pt}{\baselineskip}\selectfont \textbf{(1)}}}}{R}_{itjt}=-\frac{1}{2}\partial_{t}\partial_{t}h^{TT}_{ij}+\partial_{i}\partial_{j}\Phi +\partial_{t}\partial_{(i}\Xi_{j)}
    -\frac{1}{2}\delta_{ij}\partial_{t}\partial_{t}\Theta. \label{invariant-R}
\end{eqnarray}
Since $\Omega$, $\Psi$, and $\Sigma_i$ are not included in Eq.~\eqref{invariant-R}, they have no effect on the polarization modes of GWs. \par
According to Ref.~\cite{Eardley:1973br}, there are at most six polarization modes for a general metric theory of gravity in four dimensions. 
If taking the propagation direction of GWs as the ``$+z$" direction, then the relationship between the polarization modes and 
the components of Riemann tensor is given by~\cite{Eardley:1973br, Dong:2021jtd, Dong:2023xyb}
\begin{equation}
    \overset{\text{{\fontsize{3.5pt}{\baselineskip}\selectfont \textbf{(1)}}}}{R}_{titj}=\frac{1}{2}
    \left( 
    \begin{array}{ccccc}
        (P_b+P_+)& &P_{\times}& &P_x \\
        \\
        P_{\times}& &(P_b-P_+)& &P_y \\
        \\
        P_x& &P_y& &P_l
    \end{array}
    \right). \label{polarization-P}
\end{equation}
Here, $P_{+}$, $P_{\times}$, $P_{x}$, $P_{y}$, $P_{b}$, and $P_{l}$ are respectively the plus, cross, vector-$x$, vector-$y$, breathing, and longitudinal modes.\par
In this kind of coordinate, all gauge invariants are the functions of only $t$ and $z$. And for $\Xi_i$, $h^{TT}_{ij}$, and $\Sigma_i$, only $x$ and $y$ 
components are nonzero. By combining Eqs.~\eqref{invariant-R} and~\eqref{polarization-P}, we can obtain the relationships between gauge invariants and 
polarization modes:
\begin{equation}
    \begin{array}{lll}
        \text{tensor modes: }&P_{+}=-\partial_t\partial_t h^{TT}_{xx}=-\partial_t\partial_t h_{+},\;\;\; &P_{\times}=-\partial_t\partial_t h^{TT}_{xy}=-\partial_t\partial_t h_{\times},\\
        \text{vector modes: }&P_x=\partial_t\partial_z\Xi_x, &P_y=\partial_t\partial_z\Xi_y, \\
        \text{scalar modes: }&P_b=-\partial_t\partial_t\Theta, &P_l=2\partial_z\partial_z\Phi-\partial_t\partial_t\Theta.
    \end{array}
    \label{relationship-Pg}
\end{equation}
Clearly, the tensor perturbation $h^{TT}_{ij}$ only contributes to the tensor modes, while the vector perturbation $\Xi_i$ only contributes to the vector modes, 
and the scalar perturbations $\Theta$ and $\Phi$ only contribute to the scalar modes. 
As we are aware, the quantities ($P_i,\;i=+,\times,x,y,b,l$), representing the polarization modes, signify the relative motion between two test particles. 
The polarization modes including ``$\partial_z$'' imply their contributions in the ``$z$'' direction. Therefore, since GWs propagate along the ``$+z$'' direction, 
the three modes $P_x$, $P_y$, and $P_l$ are not transverse. \par

When we consider a specific metric theory of gravity, there are 
some constraints on these gauge invariants from field equations. Consequently, some of these gauge invariants may equal zero, and some may not be independent. 
As a result, we are left with fewer polarization modes.\par
In the following section, we will analyze the polarization modes of GWs in general Einstein-vector theory based on the method presented here.

\section{Polarization modes of GWs in general Einstein-vector theory} \label{P-modes}
In Sec.~\ref{section-Gp}, we obtained a series of gauge invariants in vector-tensor theory. Then we reviewed all possible 
polarization modes in metric theories, and derived the relationships between polarization modes and gauge invariants. Now we study 
the polarization modes of GWs in general Einstein-vector theory based on the perturbation equations~\eqref{linear1-A}-\eqref{perturbationEq-g}.
\par
According to Sec.~\ref{LPE}, the parameter spaces are limited to two cases~\eqref{case1-1627} and~\eqref{case2-1627} by the background equations~\eqref{background1-1620} and~\eqref{background2-1620}. 
\par
\textbf{For the case \ding{172}} (see Eq.~\eqref{case1-1627}): $\Lambda_0=0$, $\mathring{A}^{\mu}=0$, it is easy to find that the perturbation equation~\eqref{linear1-A} only contains the vector field, and the perturbation equation~\eqref{linear1-g} is the same as vacuum Einstein field equation. 
This implies that there is no interaction between the vector field and the gravitational field under linear approximations in the Minkowski background. 
Consequently, according to the geodesic deviation equation, the vector field has no effect on the propagation and polarizations of GWs in this case. 
Therefore, the characteristics of GWs remain the same as the case in general relativity with only two tensor modes, the plus mode and the cross mode, 
propagating at the same speed as light, see TABLE~\ref{Polarization1-table}. 
\begin{table}[h]
    \centering
\begin{tabular}{|c|c|c|c|}
    \hline
    \textbf{Conditions} & \;\textbf{Speed}\; & \;\textbf{Modes}\; & \;\textbf{DoF}\; \\ \hline
    $\Lambda_0=0$, $\mathring{A}^{\mu}=0$ & 1 & $P_+$, $P_{\times}$ & 2 \\ \hline
\end{tabular}
\caption{Polarization modes of GWs in the case of $\Lambda_0=0$ and $\mathring{A}^{\mu}=0$ for general Einstein-vector theory. The number in the last column represent the polarization degrees of freedom.}
\label{Polarization1-table}
\end{table}
\par
\textbf{For the case \ding{173}} (see Eq.~\eqref{case2-1627}): $\Lambda_0=0$, $\mu_0=0$, we will study the polarizations of GWs based on the perturbation equations~\eqref{perturbationEq-a} and~\eqref{perturbationEq-g}.
For the convenience of analysis, we will substitute Eq.~\eqref{decomposition-ha} into the perturbation equations~\eqref{perturbationEq-a} and~\eqref{perturbationEq-g}, and apply the method of gauge invariants to derive the perturbation equations consisting of gauge invariants.
\par
For the vector field equation~\eqref{perturbationEq-a}, consider the spatial and temporal components separately, and we can obtain
\begin{eqnarray}
    f^t=0 ,\qquad f^i=0.\label{ft-ei}
\end{eqnarray}
And for the metric field equation~\eqref{perturbationEq-g}, we can obtain 
\begin{eqnarray}
    \overset{\text{{\fontsize{3.5pt}{\baselineskip}\selectfont \textbf{(1)}}}}{G}_{\mu\nu}-\beta \overset{\text{{\fontsize{3.5pt}{\baselineskip}\selectfont \textbf{(1)}}}}{Y}_{\mu\nu}-\gamma \overset{\text{{\fontsize{3.5pt}{\baselineskip}\selectfont \textbf{(1)}}}}{Z}_{\mu\nu}=0. \label{Zij-ei}
\end{eqnarray}
The specific forms of $f^t$, $f^i$, $\overset{\text{{\fontsize{3.5pt}{\baselineskip}\selectfont \textbf{(1)}}}}{G}_{\mu\nu}$, $\overset{\text{{\fontsize{3.5pt}{\baselineskip}\selectfont \textbf{(1)}}}}{Y}_{\mu\nu}$, and $\overset{\text{{\fontsize{3.5pt}{\baselineskip}\selectfont \textbf{(1)}}}}{Z}_{\mu\nu}$ are listed in Appendix~\ref{Appendix1}.
\par
\indent To simplify the above equations, we can always construct a coordinate system that sets the direction of the propagation of GWs as ``$+z$'', and performs spatial rotations along the $z$-axis until $\mathring{A}^y=0$. In such coordinate, ten nontrivial gauge invariants are
$h_{+}(t,z)\equiv h^{TT}_{xx}=-h^{TT}_{yy}$, $h_{\times}(t,z)\equiv h^{TT}_{xy}$, $\Xi_x(t,z)$, $\Xi_y(t,z)$, $\Theta(t,z)$, $\Phi(t,z)$, $\Sigma_x(t,z)$, $\Sigma_y(t,z)$, $\Psi(t,z)$, and $\Omega(t,z)$. And the relationships between these gauge invariants and polarization modes are 
obtained in Eq.~\eqref{relationship-Pg}. 

\subsection{The velocities of GWs} \label{velocitySolutions}
We consider monochromatic plane waves along ``$+z$":
\begin{equation}
    \Pi(x)=\Pi(k)e^{ik_{\mu}x^{\mu}}=\Pi(k)e^{i(k_0t+k_ix^i)}=\Pi(k)e^{i(-wt+k z)}, \label{velocity-1.23.11.02}
\end{equation}
where $\Pi$ represents $h_{+}$, $h_{\times}$, $\Xi_x$, $\Xi_y$, $\Theta$, $\Phi$, $\Sigma_x$, $\Sigma_y$, $\Psi$, and $\Omega$. The velocity of plane waves is $v=\frac{\mathrm{d}w}{\mathrm{d}k}$. Substituting Eq.~\eqref{velocity-1.23.11.02} into Eqs.~\eqref{ft-ei}-\eqref{Zij-ei} and considering the coordinate that ``$\mathring{A}^y=0$", 
then setting the coefficient determinant of the equations to zero, we can obtain the velocity solutions of plane waves:
\begin{eqnarray}
    v_1&=&\pm \Big[ 1+\frac{\gamma}{2}\big(\mathring{A}^t\mp\mathring{A}^z\big)^2+\mathcal{O}\big(\big|\gamma\mathring{A}^{\mu}\mathring{A}^{\nu}\big|^2\big) \Big], \label{v1-2012}\\
    v_2&=&\pm \Big[ 1+\frac{\gamma^2}{2}\big(\mathring{A}^t\mp\mathring{A}^z\big)^2+\mathcal{O}\big(\big|\gamma^2\mathring{A}^{\mu}\mathring{A}^{\nu}\big|^2\big) \Big],\label{v2-2012}\\
    v_3&=&\pm \Big[ 1+\frac{\gamma(4\gamma-1)}{6}\big(\mathring{A}^t\mp\mathring{A}^z\big)^2+\mathcal{O}\big(\big|\gamma(4\gamma-1)\mathring{A}^{\mu}\mathring{A}^{\nu}\big|^2\big) \Big],\label{v3-2012}\\
    v_4&=&\frac{\mathring{A}^z}{\mathring{A}^t}.
\end{eqnarray}
The full forms of these velocities are represented in Appendix~\ref{Appendix2}.
The corresponding dispersion relations are $w=v_i k$ for $i=1,2,3,4$. Since we have not detected any Lorentz-violating effects in gravitational experiments, the coupling terms ($\beta R A^{2}$ and $\gamma G_{\mu\nu}A^{\mu}A^{\nu}$) in the action~\eqref{action2-2104} should be very small compared to the curvature. So it needs to satisfy $|\beta\mathring{A}^{\mu}\mathring{A}^{\nu}|\ll 1$, and $|\gamma\mathring{A}^{\mu}\mathring{A}^{\nu}|\ll 1$. 
Then we can expand the velocities like above.\par

For the velocity ``$v=v_4$", all gauge invariants ($h_{+}, h_{\times}, \Xi_x, \Xi_y, \Theta, \Phi$) that contribute to GWs are 0.
So ``$v=v_4$" is not a solution for the velocity of GWs.\par
For the velocities ``$v=v_1$", ``$v=v_2$", and ``$v=v_3$", if we denote ``$v_+$'' and ``$v_-$'' respectively for the positive and negative solutions, then they represent GWs traveling along the ``$+z$" and ``$-z$'' directions, respectively. And $v_+$ and $v_-$ have the following relation:
\begin{equation}
    v_+(\mathring{A}^z)=-v_-(-\mathring{A}^z).
\end{equation}
This relation represents the operation $\hat{P}$, rotating the spatial coordinate $180$ degrees along the $x$-axis, which satisfies $\hat{P}(v_-)=v_+$ and $\hat{P}(v_+)=v_-$. It means physically $v_+$ and $v_-$ are the same velocity solutions, just the results in different coordinate systems.
Therefore, it is sufficient to consider only the positive velocity solutions ``$v_+$" (propagating along ``$+z$''). 
Then according to Eqs.~\eqref{v1-2012} and~\eqref{v2-2012}, it is easy to find $v_1=v_2=v_3=1$ when $\mathring{A}^t=\mathring{A}^z$.
\par

Under the special coordinate that used above, the field equations consisting of gauge invariants, Eqs.~\eqref{ft-ei}-\eqref{Zij-ei}, can be divided into two groups. The first group, consisting of four equations, involves only $h_{\times}$, $\Xi_y$, and $\Sigma_y$. 
The second group, consisting of the other ten equations, involves only $h_+$, $\Xi_x$, $\Theta$, $\Phi$, $\Sigma_x$, $\Psi$, and $\Omega$.
We will analyze the two groups of equations individually within such coordinate.

\subsection{First group: $h_{\times}, \Xi_y, \Sigma_y$}
In the coordinate that GWs propagate along ``$+z$'' and ``$\mathring{A}^y=0$'', there are four equations involving and only involving gauge invariants $h_{\times}$, $\Xi_y$, and $\Sigma_y$ in Eqs.~\eqref{ft-ei}-\eqref{Zij-ei}. They are 
\begin{eqnarray}
    f^y&=&0,\label{equations-Fg1}\\
    F_{ty}&\equiv& \overset{\text{{\fontsize{3.5pt}{\baselineskip}\selectfont \textbf{(1)}}}}{G}_{ty}-\beta \overset{\text{{\fontsize{3.5pt}{\baselineskip}\selectfont \textbf{(1)}}}}{Y}_{ty}-\gamma \overset{\text{{\fontsize{3.5pt}{\baselineskip}\selectfont \textbf{(1)}}}}{Z}_{ty}=0,\\
    F_{xy}&\equiv& \overset{\text{{\fontsize{3.5pt}{\baselineskip}\selectfont \textbf{(1)}}}}{G}_{xy}-\beta \overset{\text{{\fontsize{3.5pt}{\baselineskip}\selectfont \textbf{(1)}}}}{Y}_{xy}-\gamma \overset{\text{{\fontsize{3.5pt}{\baselineskip}\selectfont \textbf{(1)}}}}{Z}_{xy}=0,\\
    F_{yz}&\equiv& \overset{\text{{\fontsize{3.5pt}{\baselineskip}\selectfont \textbf{(1)}}}}{G}_{yz}-\beta \overset{\text{{\fontsize{3.5pt}{\baselineskip}\selectfont \textbf{(1)}}}}{Y}_{yz}-\gamma \overset{\text{{\fontsize{3.5pt}{\baselineskip}\selectfont \textbf{(1)}}}}{Z}_{yz}=0.\label{equations-Fg4}
\end{eqnarray}
Here, the specific forms of $f^y,\;F_{ty},\;F_{xy}$ and $F_{yz}$ are given in Appendix~\ref{Appendix3}. 
There are four equations involving only three variables, which may lead to no solution of the equations. However, if we consider monochromatic plane GWs, then the equation $F_{yz}=0$ is equivalent to the equation $F_{ty}=0$. Since the detectors are far enough from the wave source, this ansatz of monochromatic plane GWs is reasonable.\par

According to Eqs.~\eqref{equations-Fg1}-\eqref{equations-Fg4} and considering monochromatic plane GWs, we can analyze  the polarization modes of GWs associated with the gauge invariants $h_{\times}$, $\Xi_y$, and $\Sigma_y$. Then we  find the polarization modes of GWs depend on the parameter spaces of $\gamma$, $\mathring{A}^x$, $\mathring{A}^z$, and $\mathring{A}^t$. 
By considering all situations, we can categorize them into the following three cases.\\

\textbf{Case $\mathcal{I}$:} ``$\gamma\ne 0,1; \mathring{A}^x\ne 0; \mathring{A}^t\ne 0 \;\text{or}\; \mathring{A}^z\ne 0; \mathring{A}^t\ne\mathring{A}^z$''.\par
In this case, by solving Eqs.~\eqref{equations-Fg1}-\eqref{equations-Fg4} and considering the polarization relations~\eqref{relationship-Pg}, we obtain two groups of solutions: \\
\indent Solution \ding{172}:
\begin{eqnarray}
    && v=v_1\approx 1+\frac{\gamma}{2}\big(\mathring{A}^t-\mathring{A}^z\big)^2, \qquad P_y =\frac{(v^2-1)\mathring{A}^x}{v(v\mathring{A}^z-\mathring{A}^t)}P_{\times},  \label{ymode-yh1}
    \\
    &&\Big[ \big(2 +D_1(\mathring{A}^t,\mathring{A}^x,\mathring{A}^z)\big)\partial_{z}\partial_{z}+4\gamma\mathring{A}^{t}\mathring{A}^{z}\partial_{t}\partial_{z}-\big(2 -D_2(\mathring{A}^t,\mathring{A}^x,\mathring{A}^z)\big)\partial_{t}\partial_{t} \Big] \Gamma=0 \label{wave-hx1}.
\end{eqnarray}
\indent Solution \ding{173}: 
\begin{eqnarray}
    && v=v_2\approx 1+\frac{\gamma^2}{2}\big(\mathring{A}^t-\mathring{A}^z\big)^2,\qquad P_y =-\frac{v\mathring{A}^z-\mathring{A}^t}{v\mathring{A}^x}P_{\times},  \label{ymode-yh2}
    \\
    &&\Big[ \big(2 +D_3(\mathring{A}^t,\mathring{A}^x,\mathring{A}^z)\big)\partial_z\partial_z+4\gamma^{2}\mathring{A}^{t}\mathring{A}^{z}\partial_{t}\partial_{z}-\big(2 -D_4(\mathring{A}^t,\mathring{A}^x,\mathring{A}^z) \big)\partial_{t}\partial_{t} \Big]\Gamma=0. \label{wave-hx2}
\end{eqnarray}
The specific forms of $D_1,\;D_2,\;D_3,\;D_4,\;D_5,$ and $D_6$ used here and below are given in Appendix~\ref{Appendix3}.
Here, Eqs.~\eqref{wave-hx1} and~\eqref{wave-hx2} are the wave equations of the $P_{\times}$ and $P_y$ modes, and $\Gamma$ stands for $h_{\times}$ or $\Xi_y$. Equations~\eqref{ymode-yh1} and~\eqref{ymode-yh2} are constraints between 
polarization modes.  
Since $w,k\ge 0$, so $v\ge 0$ (see Sec.~\ref{velocitySolutions}), we don't consider the solutions that the velocity of GWs is negative in this and the following calculations. \par

It is easy to see that the $P_y$ mode depends on the $P_{\times}$ mode in both solutions.
Since no Lorentz-violating effects have been detected in gravitational experiments, the coupling terms in the action should be very small and satisfy $|\beta\mathring{A}^{\mu}\mathring{A}^{\nu}|\ll 1$, $|\gamma\mathring{A}^{\mu}\mathring{A}^{\nu}|\ll 1$. Based on this, we can analyze the order of magnitude of the polarization modes:
\begin{eqnarray}
    \text{Solution \ding{172}:}
    &&P_y=\frac{(v^2-1)\mathring{A}^x}{v(v\mathring{A}^z-\mathring{A}^t)}P_{\times}=\mathcal{O}(|\gamma\mathring{A}^{\mu}\mathring{A}^{\nu}|)P_{\times}.\label{solution-1}\\
    \text{Solution \ding{173}:}
    &&P_y=-\frac{v\mathring{A}^z-\mathring{A}^t}{v\mathring{A}^x}P_{\times}=\mathcal{O}(1)P_{\times}.\label{solution-2}
\end{eqnarray}
By combining the two situations, we can observe that there are two mixture modes of the cross mode ($P_{\times}$) and the vector-$y$ mode ($P_y$) with different speeds in this case. In one mixture mode, the amplitude of $P_y$ is suppressed by that of $P_{\times}$. In another mixture mode, the amplitudes of $P_y$ and $P_{\times}$ are the same order of magnitude. And the $\Sigma_y$ depends on the cross mode, but it does not contribute to GWs.\\

\textbf{Case $\mathcal{II}$:} ``$\gamma\ne 0; \gamma=1 \;\text{or}\; \mathring{A}^x=0; \mathring{A}^t\ne 0\; \text{or}\; \mathring{A}^z\ne 0; \mathring{A}^t\ne\mathring{A}^z$''.\par
In this case, by solving Eqs.~\eqref{equations-Fg1}-\eqref{equations-Fg4} and considering the polarization relations~\eqref{relationship-Pg}, we obtain two groups of solutions:\\
\indent Solution \ding{172}:
\begin{eqnarray}
    && v=v_1\approx 1+\frac{\gamma}{2}\big(\mathring{A}^t-\mathring{A}^z\big)^2 ,\qquad P_y=0,
    \\
    &&\Big[ \big(2 +D_1(\mathring{A}^t,\mathring{A}^z)  \big)\partial_{z}\partial_{z} +4\gamma\mathring{A}^{t}\mathring{A}^{z}\partial_{t}\partial_{z}-\big(2 -D_2(\mathring{A}^t,\mathring{A}^z) \big)\partial_{t}\partial_{t} \Big] h_{\times}=0. \label{wave-htimes1}
\end{eqnarray}
\indent Solution \ding{173}:
\begin{eqnarray}
    && v=v_2\approx 1+\frac{\gamma^2}{2}\big(\mathring{A}^t-\mathring{A}^z\big)^2 ,\qquad P_{\times}=0,
    \\
    &&\Big[ \big(2 +D_3(\mathring{A}^t,\mathring{A}^z)  \big)\partial_{z}\partial_{z} +4\gamma^2\mathring{A}^{t}\mathring{A}^{z}\partial_{t}\partial_{z}-\big(2 -D_4(\mathring{A}^t,\mathring{A}^z) \big)\partial_{t}\partial_{t} \Big]\Xi_y=0.\label{wave-hy1141}
\end{eqnarray}
Here we just present a set of representative solutions (the solutions with $\mathring{A}^x=0$), and the solutions with $\mathring{A}^x\ne 0$ in this case deffer only in coefficient compared to the above solutions. According to $P_{\times}=-\partial_t\partial_t h_{\times}$ and $P_y=\partial_t\partial_z\Xi_y$, there are two independent polarization modes $P_{\times}$ and $P_y$ in this case. When $\gamma=1$, the two modes of GWs have the same speed $v_1=v_2$.\\

\textbf{Case $\mathcal{III}$:} ``$\gamma=0 \;\text{or}\; \mathring{A}^t=\mathring{A}^z$''.\par
For this case, we get the following simple result:
\begin{equation}
    \begin{split}
        &v=1,\quad P_y=0,\quad
        (\partial_t\partial_t-\partial_z\partial_z)h_{\times}=0.
    \end{split}
\end{equation}
According to $P_{\times}=-\partial_t\partial_t h_{\times}$, there is only one $P_{\times}$ mode with the same speed as light in this case.\\

Based on the above research, we summarize the results in TABLE~\ref{Polarization2-table}. 
The first column of the table lists the conditions for the parameters $\gamma$, $\mathring{A}^t$, $\mathring{A}^x$, and $\mathring{A}^z$. The second column provides the speed of GWs. 
The third column presents the constraint between the $P_{\times}$ and $P_{y}$ modes. The fourth column lists the all possible polarization modes. Finally, The last column gives the polarization degrees of freedom.
\begin{table}[h]
    \centering
    \begin{footnotesize}
    \begin{tabular}{|l|c|l|l|c|}
        \hline
        \textbf{Conditions} & \textbf{Speed} & \textbf{Constraint} & \textbf{Modes} & \textbf{DoF}\\ \hline
        \multirow{2}*{Case $\mathcal{I}$: $\gamma\ne 0,1; \; \mathring{A}^x\ne 0; \; \mathring{A}^t\ne 0 \;\text{or}\; \mathring{A}^z\ne 0; \; \mathring{A}^t\ne\mathring{A}^z$.} &
            $v_1$ & $P_y=\frac{(v^2-1)\mathring{A}^x}{v(v\mathring{A}^z-\mathring{A}^t)}P_{\times}$ & $P_{\times}$, $P_y$ & 1 \\ \cline{2-5}
            & $v_2$ & $P_y=-\frac{v\mathring{A}^z-\mathring{A}^t}{v\mathring{A}^x}P_{\times}$ & $P_{\times}$, $P_y$ & 1 \\ \hline
        \multirow{2}*{Case $\mathcal{II}$: $\gamma\ne 0; \; \gamma=1 \;\text{or}\; \mathring{A}^x=0; \; \mathring{A}^t\ne 0\; \text{or}\; \mathring{A}^z\ne 0; \; \mathring{A}^t\ne\mathring{A}^z$.} &
            $v_1$ & & $P_{\times}$ & 1 \\ \cline{2-5}
            & $v_2$ & & $P_y$ & 1 \\ \hline
        Case $\mathcal{III}$: $\gamma=0 \;\text{or}\; \mathring{A}^t=\mathring{A}^z$. & 1 & & $P_{\times}$ & 1 \\ \hline
    \end{tabular}
\end{footnotesize}
\caption{Polarization modes of GWs for the first group ($h_{\times}, \Xi_y, \Sigma_y$) in the case of $\Lambda_0=0$ and $\mu_0=0$ for general Einstein-vector theory.}
\label{Polarization2-table}
\end{table}

\subsection{Second group: $h_+, \Xi_x, \Theta, \Phi, \Sigma_x, \Omega, \Psi$} \label{secondgroup-1544}
At the beginning of this section, we have derived fourteen field equations involving the gauge invariants, as shown in Eqs.~\eqref{ft-ei}-\eqref{Zij-ei}. In the coordinate where GWs propagate along the ``$+z$'' and ``$\mathring{A}^y=0$'', we have analyzed four of them which involve only $h_{\times}$, $\Xi_y$, and $\Sigma_y$.
Now we will analyze the remaining ten equations which involve only $h_+$, $\Xi_x$, $\Theta$, $\Phi$, $\Sigma_x$, $\Omega$, and $\Psi$ in the same coordinate.
The ten equations are listed as follows:
\begin{equation}
    \begin{split}
        &f^t=0,\quad f^x=0,\quad f^z=0,\quad \overset{\text{{\fontsize{3.5pt}{\baselineskip}\selectfont \textbf{(1)}}}}{G}_{tt}=\gamma \overset{\text{{\fontsize{3.5pt}{\baselineskip}\selectfont \textbf{(1)}}}}{Y}_{tt}+\beta \overset{\text{{\fontsize{3.5pt}{\baselineskip}\selectfont \textbf{(1)}}}}{Z}_{tt},\quad \overset{\text{{\fontsize{3.5pt}{\baselineskip}\selectfont \textbf{(1)}}}}{G}_{tx}=\gamma \overset{\text{{\fontsize{3.5pt}{\baselineskip}\selectfont \textbf{(1)}}}}{Y}_{tx}+\beta \overset{\text{{\fontsize{3.5pt}{\baselineskip}\selectfont \textbf{(1)}}}}{Z}_{tx},\quad \overset{\text{{\fontsize{3.5pt}{\baselineskip}\selectfont \textbf{(1)}}}}{G}_{tz}=\gamma \overset{\text{{\fontsize{3.5pt}{\baselineskip}\selectfont \textbf{(1)}}}}{Y}_{tz}+\beta \overset{\text{{\fontsize{3.5pt}{\baselineskip}\selectfont \textbf{(1)}}}}{Z}_{tz},\\
        &\overset{\text{{\fontsize{3.5pt}{\baselineskip}\selectfont \textbf{(1)}}}}{G}_{xx}=\gamma \overset{\text{{\fontsize{3.5pt}{\baselineskip}\selectfont \textbf{(1)}}}}{Y}_{xx}+\beta \overset{\text{{\fontsize{3.5pt}{\baselineskip}\selectfont \textbf{(1)}}}}{Z}_{xx},\quad\; \overset{\text{{\fontsize{3.5pt}{\baselineskip}\selectfont \textbf{(1)}}}}{G}_{xz}=\gamma \overset{\text{{\fontsize{3.5pt}{\baselineskip}\selectfont \textbf{(1)}}}}{Y}_{xz}+\beta \overset{\text{{\fontsize{3.5pt}{\baselineskip}\selectfont \textbf{(1)}}}}{Z}_{xz},\quad\; \overset{\text{{\fontsize{3.5pt}{\baselineskip}\selectfont \textbf{(1)}}}}{G}_{yy}=\gamma \overset{\text{{\fontsize{3.5pt}{\baselineskip}\selectfont \textbf{(1)}}}}{Y}_{yy}+\beta \overset{\text{{\fontsize{3.5pt}{\baselineskip}\selectfont \textbf{(1)}}}}{Z}_{yy},\quad\;
        \overset{\text{{\fontsize{3.5pt}{\baselineskip}\selectfont \textbf{(1)}}}}{G}_{zz}=\gamma \overset{\text{{\fontsize{3.5pt}{\baselineskip}\selectfont \textbf{(1)}}}}{Y}_{zz}+\beta \overset{\text{{\fontsize{3.5pt}{\baselineskip}\selectfont \textbf{(1)}}}}{Z}_{zz}.
    \end{split}\label{fieldequation-all}
\end{equation}
By considering monochromatic plane GWs propagating along ``$+z$'' and using Eqs.~\eqref{fieldequation-all}, we can analyze $h_+$, $\Xi_x$, $\Theta$, $\Phi$, $\Sigma_x$, $\Omega$, and $\Psi$, as well as the polarization modes of GWs involving these gauge invariants. We find that the polarization modes of GWs depend
on the parameter spaces of $\beta$, $\gamma$, $\mathring{A}^x$, $\mathring{A}^z$, and $\mathring{A}^t$. By combining all situations, we can identify the following five cases.\\

\textbf{Case $\mathcal{I}$:} ``$\gamma\ne 0,1; \mathring{A}^x\ne 0; \mathring{A}^t\ne 0\; \text{or}\; \mathring{A}^z\ne 0; \mathring{A}^t\ne \mathring{A}^z$''.\par
In this case, substituting these restrictions into Eq.~\eqref{fieldequation-all}, solving these equations and considering the polarization relations~\eqref{relationship-Pg}, we can obtain the following solutions:\\
\indent Solution \ding{172}:
\begin{eqnarray}
    && v=v_1\approx 1+\frac{\gamma}{2}\big(\mathring{A}^t-\mathring{A}^z\big)^2,\\
    && P_l=2\Big(\frac{1}{v^2}-1 \Big)P_b, \quad P_x=\frac{2(\mathring{A}^t-v\mathring{A}^z)}{v\mathring{A}^x}P_b,\quad P_b=-\frac{\gamma\mathring{A}^x\mathring{A}^x}{2+K_1}P_+,\label{11wave3-1801}
    \\
    && \Big[ \big(2+D_1(\mathring{A}^t,\mathring{A}^x,\mathring{A}^z)\big)\partial_z\partial_z + 4\gamma\mathring{A}^t\mathring{A}^z\partial_z\partial_t- \big(2-D_2(\mathring{A}^t,\mathring{A}^x,\mathring{A}^z)\big)\partial_t\partial_t \Big]\Upsilon_4=0.\label{upsilon1-1742}
\end{eqnarray}
\indent Solution \ding{173}:
\begin{eqnarray}
    && v=v_2\approx 1 +\frac{\gamma^2}{2}\big(\mathring{A}^t-\mathring{A}^z\big)^2,\\
    && P_l=2\Big(\frac{1}{v^2}-1\Big)P_b,\quad P_b=P_+,\quad P_x=\frac{2+K_2}{2+K_3}\frac{\big(\mathring{A}^t-v\mathring{A}^z\big)}{v\mathring{A}^x}P_b, \label{12wave3-1801}
    \\
    && \Big[ \big(2+D_3(\mathring{A}^t,\mathring{A}^x,\mathring{A}^z)\big)\partial_z\partial_z +4\gamma^2\mathring{A}^t\mathring{A}^z\partial_z\partial_t-\big(2-D_4(\mathring{A}^t,\mathring{A}^x,\mathring{A}^z)\big)\partial_t\partial_t \Big] \Upsilon_4=0.\label{upsilon2-1742}
\end{eqnarray}
\indent Solution \ding{174}:
\begin{eqnarray}
    && v=v_3\approx 1 +\frac{\gamma (4\gamma-1)}{6}\big(\mathring{A}^t-\mathring{A}^z\big)^2,\\
    && P_l=2\Big(\frac{1}{v^2}-1\Big)P_b,\qquad P_x=\frac{2(\mathring{A}^t-v\mathring{A}^z)}{v\mathring{A}^x}P_+,\quad P_+= -\frac{\gamma(4\gamma-1)\mathring{A}^x\mathring{A}^x}{2+K_2}P_b,\label{13wave3-1801}
    \\
    && \Big[ \big(6+D_5(\mathring{A}^t,\mathring{A}^x,\mathring{A}^z)\big)\partial_z\partial_z +4\gamma(4\gamma-1)\mathring{A}^t\mathring{A}^z\partial_z\partial_t-\big(6- D_6(\mathring{A}^t,\mathring{A}^x,\mathring{A}^z)\big)\partial_t\partial_t \Big]\Upsilon_4=0.\label{upsilon3-1742}
\end{eqnarray}
Here $K_i=c_{it}\mathring{A}^t\mathring{A}^t+c_{ix}\mathring{A}^x\mathring{A}^x+c_{iz}\mathring{A}^z\mathring{A}^z\; (i=1,2,3)$, $c_{it},c_{ix},c_{iz}\text{are constants}$.
We just present a set of representative solutions (the solutions with $\beta\ne 0$ and $\gamma\ne \frac{1}{4}$). Equations~\eqref{upsilon1-1742},~\eqref{upsilon2-1742}, and~\eqref{upsilon3-1742} are the wave equations for polarization modes of GWs, while $\Upsilon_4$ stands for $h_+$, $\Xi_x$, $\Theta$, or $\Phi$. Equations~\eqref{11wave3-1801},~\eqref{12wave3-1801}, and~\eqref{13wave3-1801} are the constraint equations between the amplitudes of polarization modes of GWs. Obviously, the $P_b$, $P_l$, and $P_x$ modes all depend on $P_+$ mode in both solutions \ding{172} and \ding{173}. For solution \ding{174}, the $P_+$, $P_l$, and $P_x$ modes all depend on the $P_b$ mode. If adding $\beta= 0$ in this case and solving equations~\eqref{fieldequation-all}, one can find there is no mode allowed in solution \ding{174}. Furthermore, it is easy to observe that there is only $P_b$ mode in solution \ding{174} when adding $\beta\ne 0$ and $\gamma= \frac{1}{4}$ in this case.\par
According to these constraint equations, we can analyze the relations of amplitudes of the polarization modes:
\begin{eqnarray}
    \text{Solution \ding{172}:}&& P_b=\mathcal{O}\big(\big|\gamma\mathring{A}^{\mu}\mathring{A}^{\nu}\big|\big)P_+,\quad P_l= \mathcal{O}\big(\big|\gamma\mathring{A}^{\mu}\mathring{A}^{\nu}\big|^2\big)P_+,\quad P_x=\mathcal{O}\big(\big|\gamma\mathring{A}^{\mu}\mathring{A}^{\nu}\big|\big)P_+.\\
    \text{Solution \ding{173}:}&& P_b=\mathcal{O}(1)P_+,\qquad P_l= \mathcal{O}\big(\big|\gamma^2\mathring{A}^{\mu}\mathring{A}^{\nu}\big|\big)P_+,\qquad P_x=\mathcal{O}(1)P_+.\\
    \text{Solution \ding{174}:}&& P_b=\mathcal{O}\big(\big|\gamma(4\gamma-1)\mathring{A}^{\mu}\mathring{A}^{\nu}\big|^{-1}\big)P_+,\quad P_l=\mathcal{O}(1)P_+,\quad P_x=\mathcal{O}(1)P_+.
\end{eqnarray}
Through the above analysis, we have presented the relationships between the amplitudes of $P_b$, $P_l$, and $P_x$ and that of $P_+$ . The amplitudes of $P_b$, $P_l$, and $P_x$ compared to that of $P_+$ are different in different solutions. In particular, the amplitude of $P_b$ is significantly larger than that of $P_+$ in solution \ding{174}.\\

\textbf{Case $\mathcal{II}$:} ``$\gamma \ne 0; \mathring{A}^x=0; \mathring{A}^t\ne 0 \;\text{or}\; \mathring{A}^z\ne 0;  \mathring{A}^t\ne \mathring{A}^z$'' or ``$\beta\ne 0; \gamma=1; \mathring{A}^x\ne 0; \mathring{A}^t\ne 0 \;\text{or}\; \mathring{A}^z\ne 0;  \mathring{A}^t\ne \mathring{A}^z$''. \par
Substituting these restrictions into Eq.~\eqref{fieldequation-all}, solving these equations and considering the polarization relations~\eqref{relationship-Pg}, we get the following solutions:\\
\indent Solution \ding{172}:
\begin{eqnarray}
    && v=v_1\approx 1+\frac{\gamma}{2}\big(\mathring{A}^t-\mathring{A}^z\big)^2 ,\qquad P_l=P_b=P_x=0,
    \\
    && \Big[ \big(2+D_1(\mathring{A}^t,\mathring{A}^z)\big)\partial_z\partial_z+ 4\gamma\mathring{A}^t\mathring{A}^z\partial_z\partial_t - \big(2-D_2(\mathring{A}^t,\mathring{A}^z)\big)\partial_t\partial_t \Big]h_+=0.\label{wave1-1754}
\end{eqnarray}
\indent Solution \ding{173}:
\begin{eqnarray}
    && v=v_2\approx 1 +\frac{\gamma^2}{2}\big(\mathring{A}^t-\mathring{A}^z\big)^2 ,\qquad P_l=P_b=P_+=0,
    \\
    && \Big[ \big(2+D_3(\mathring{A}^t,\mathring{A}^z)\big)\partial_z\partial_z +4\gamma^2\mathring{A}^t\mathring{A}^z\partial_z\partial_t -\big(2-D_4(\mathring{A}^t,\mathring{A}^z)\big)\partial_t\partial_t \Big]\Xi_x=0.\label{wave2-1754}
\end{eqnarray}
\indent Solution \ding{174}:
\begin{eqnarray}
    && v=v_3\approx 1 +\frac{\gamma (4\gamma-1)}{6}\big(\mathring{A}^t-\mathring{A}^z\big)^2,\qquad P_l=2\Big(\frac{1}{v^2}-1\Big)P_b,\qquad P_x=P_+=0,
    \\
    && \Big[ \big(6+D_5(\mathring{A}^t,\mathring{A}^z)\big)\partial_z\partial_z +4\gamma(4\gamma-1)\mathring{A}^t\mathring{A}^z\partial_z\partial_t -\big(6-D_6(\mathring{A}^t,\mathring{A}^z)\big)\partial_t\partial_t \Big]\Upsilon_2=0.\label{wave3-1754}
\end{eqnarray}
We just present a set of representative solutions (the solutions with $\beta\ne 0$ and $\mathring{A}^x=0$. Notion: if $\beta=0$, there will be no solution \ding{174}). Equations~\eqref{wave1-1754},~\eqref{wave2-1754}, and~\eqref{wave3-1754} are the wave equations for polarization modes of GWs, while $\Upsilon_2$ stands for $\Theta$ or $\Phi$.  According to the above constraint equations, we can see that the $P_l$ mode depends on the $P_b$ mode, and the relation between their amplitudes is $P_l=\mathcal{O}(|\gamma(4\gamma-1)\mathring{A}^{\mu}\mathring{A}^{\nu}|)P_b$. If $\gamma=1$, the $P_+$ mode, the $P_x$ mode, and the mixture of the $P_l$ and $P_b$ modes have the same speed. For the solution \ding{174}, it is easy to find that the $P_l$ mode will vanish when $\gamma=\frac{1}{4}$.\par
In this case, the $P_+$ and $P_x$ modes are independent. And when $\beta\ne 0$, there is another $P_b$ mode ($\gamma=\frac{1}{4}$) or a mixture of the $P_b$ and $P_l$ modes ($\gamma\ne \frac{1}{4}$).\\

\textbf{Case $\mathcal{III}$:} ``$\beta=0; \gamma= 1;\;\mathring{A}^x\ne 0; \mathring{A}^t\ne 0\; \text{or}\; \mathring{A}^z\ne 0; \mathring{A}^t\ne \mathring{A}^z$''. \par
For this case, it can be shown that there is only one solution:
\begin{eqnarray}
    && v=v_1=v_2=v_3\approx 1+\frac{1}{2}\big(\mathring{A}^t-\mathring{A}^z\big)^2,\\
    &&  P_l=2\Big(\frac{1}{v^2}-1\Big)P_b,\quad \mathring{A}^xv P_x - \frac{2+K_4}{2+K_5}\big(\mathring{A}^t -\mathring{A}^z v\big)P_b +\frac{\mathring{A}^x\mathring{A}^x\big(\mathring{A}^t-\mathring{A}^z v\big)}{2+K_5}P_+=0,\label{pxmode1202}\\
    && \Big[ \big(2+\mathring{A}^t\mathring{A}^t-\mathring{A}^x\mathring{A}^x+\mathring{A}^z\mathring{A}^z\big)\partial_z\partial_z -\big(2-\mathring{A}^t\mathring{A}^t-\mathring{A}^x\mathring{A}^x-\mathring{A}^z\mathring{A}^z\big)\partial_t\partial_t +4\mathring{A}^t\mathring{A}^z\partial_t\partial_z \Big]\Upsilon_4=0. \label{upsilon-waveequation1200}
\end{eqnarray}
Here $K_i=c_{it}\mathring{A}^t\mathring{A}^t+c_{ix}\mathring{A}^x\mathring{A}^x+c_{iz}\mathring{A}^z\mathring{A}^z\; (i=4,5)$, $c_{it},c_{ix},c_{iz}\text{are constants}$, 
and $\Upsilon_4$ stands for $h_+$, $\Xi_x$, $\Theta$, or $\Phi$. Equation~\eqref{upsilon-waveequation1200} is wave equation of GWs. Equations~\eqref{pxmode1202} are the constraint equations between the amplitudes of polarization modes of GWs.
According to Eqs.~\eqref{pxmode1202}-\eqref{upsilon-waveequation1200}, there are four polarization modes in this case, but only two of them are independent. Since the tensor modes have been detected, at least one of the $P_b$ and $P_x$ modes exists by analyzing the constraint~\eqref{pxmode1202}. And the $P_l$ mode depends on the $P_b$ one. 
 \par
Then according to the constraint equations~\eqref{pxmode1202}, we can analyze the relations of amplitudes of the polarization modes:
\begin{equation}
    P_l= \mathcal{O}\big(\big|\mathring{A}^{\mu}\mathring{A}^{\nu}\big|\big)P_b,\quad P_x= \mathcal{O}(1)P_b-\mathcal{O}\big(\big|\mathring{A}^{\mu}\mathring{A}^{\nu}\big|\big)P_+.
\end{equation}
According to the above analysis about the amplitudes of the polarization modes, we can observe that the $P_l$ mode is suppressed compared to the $P_b$ one. But for the $P_+$, $P_b$, and $P_x$ modes, since there are three modes with only one constraint equation~\eqref{pxmode1202}, we can not provide specific relations for the magnitudes of any two amplitudes. However, if we assume $\mathcal{O}(P_b) \lesssim \mathcal{O}(P_+)$, then we can obtain the coarse order of the magnitude of the $P_x$ mode:
\begin{eqnarray}
    \mathcal{O}(P_b) \;\lesssim\; \mathcal{O}(P_x) \;\lesssim\; \mathcal{O}(P_+). 
\end{eqnarray}
And if we assume $\mathcal{O}(P_b) \sim \mathcal{O}(P_+)$, we have
\begin{eqnarray}
    \mathcal{O}(P_b) \;\sim\; \mathcal{O}(P_x) \;\sim\; \mathcal{O}(P_+). 
\end{eqnarray}

\indent \textbf{Case $\mathcal{IV}$:} ``$\beta\ne \frac{1}{2}\gamma; \mathring{A}^x\ne 0; \mathring{A}^t=\mathring{A}^z$'' or ``$\beta\ne 0; \gamma=0; \mathring{A}^x\ne 0\; \text{or}\; \mathring{A}^t\ne 0\; \text{or}\; \mathring{A}^z\ne 0$''.\par
For this case, we get the following simple result:
\begin{equation}
    v=1,\quad P_l=P_x=0,\quad (\partial_t\partial_t-\partial_z\partial_z)h_+=0,\quad (\partial_t\partial_t-\partial_z\partial_z)\Theta=0.
\end{equation}
If $\mathring{A}^x\ne 0$, there is another constraint: $2\mathring{A}^x\partial_t\partial_t\Sigma_x- \mathring{A}^x\mathring{A}^x P_+ -2\big(\mathring{A}^x\mathring{A}^x-1\big)P_b=0$. Since $\Sigma_x$ is gauge invariant and does not contribute to GWs, this equation does not provide a constraint on the polarization modes of GWs.
Therefore, there are only $P_+$ and $P_b$ modes with the same speed as light in this case.\\

\textbf{Case $\mathcal{V}$:} ``$\beta=\gamma=0$'' or ``$\beta=\frac{1}{2}\gamma \;\text{or}\; \mathring{A}^x=0; \mathring{A}^t=\mathring{A}^z$''. \par
In this case, we reach a simpler result:
\begin{equation}
    v=1,\quad P_l=P_b=P_x=0,\quad (\partial_t\partial_t-\partial_z\partial_z)h_+=0.
\end{equation}
Obviously, there is only $h_+$ mode in this case with the same speed as light. \\

Based on the research presented above, we summarize the results in TABLE~\ref{Polarization3-table}. 
The first column of the table lists the conditions for the parameters $\beta$, $\gamma$, $\mathring{A}^t$, $\mathring{A}^x$, and $\mathring{A}^z$. The second column provides the speed of GWs. 
The third column presents the constraints between the $P_+$, $P_{x}$, $P_b$, and $P_l$ modes. The fourth column lists the possible polarization modes. And the last column gives the polarization degrees of freedom. 
\begin{table}[h]
    \centering
    \begin{footnotesize}
    \begin{tabular}{|l|c|l|l|c|}\hline
        \textbf{Conditions} & \textbf{Speed} & \textbf{Constraints} & \textbf{Modes} & \textbf{DoF} 
        \\ \hline
        \multirow{3}*{\makecell[l]{Case $\mathcal{I}$: \\  $\gamma\ne 0,1$; $\mathring{A}^x\ne 0$; $\mathring{A}^t\ne 0$ or $\mathring{A}^z\ne 0$; $\mathring{A}^t\ne \mathring{A}^z$.}}
            & $v_1$ & \makecell[l]{$P_b\approx -\frac{\gamma \mathring{A}^x\mathring{A}^x}{2}P_+$,\\ $P_x=\frac{2(\mathring{A}^t-v\mathring{A}^z)}{v\mathring{A}^x}P_b$,\\ $P_l=2(\frac{1}{v^2}-1)P_b$.} & $P_+$, $P_x$, $P_b$, $P_l$. & 1
            \\ \cline{2-5}
            & $v_2$ & \makecell[l]{$P_b=P_+$,\\ $P_x\approx\frac{\mathring{A}^t-v\mathring{A}^z}{v\mathring{A}^x}P_b$,\\ $P_l=2(\frac{1}{v^2}-1)P_b$.} & $P_+$, $P_x$, $P_b$, $P_l$. & 1
            \\ \cline{2-5}
            & $v_3$ & \makecell[l]{$P_+\approx -\frac{\gamma(4\gamma-1)\mathring{A}^x\mathring{A}^x}{2} P_b$,\\ $P_x=\frac{2(\mathring{A}^t-v\mathring{A}^z)}{v\mathring{A}^x}P_+$,\\ $P_l=2(\frac{1}{v^2}-1)P_b$.} & $P_+$, $P_x$, $P_b$, $P_l$. & 1
        \\ \hline
        \multirow{6}*{\makecell[l]{Case $\mathcal{II}$:\\ $\gamma\ne 0$; $\mathring{A}^x=0$; $\mathring{A}^t\ne \mathring{A}^z$; $\mathring{A}^t\ne 0$ or $\mathring{A}^z\ne 0$.\\ \;\; or\\ $\beta\ne 0$; $\gamma=1$; $\mathring{A}^x\ne 0$; $\mathring{A}^t\ne 0$ or $\mathring{A}^z\ne 0$; $\mathring{A}^t\ne \mathring{A}^z$.}}
            & \multirow{2}*{$v_1$} & \multirow{2}*{ } & \multirow{2}*{$P_+$.} & \multirow{2}*{1}\\
            &  & & &
            \\ \cline{2-5}
            & \multirow{2}*{$v_2$} & \multirow{2}*{ } & \multirow{2}*{$P_x$.} & \multirow{2}*{1}\\
            &  & & &
            \\ \cline{2-5}
            & \multirow{2}*{$v_3$} & \multirow{2}*{$P_l=2(\frac{1}{v^2}-1)P_b$.} & \multirow{2}*{$P_b$, $P_l$.} & \multirow{2}*{1}\\
            &  & & &
        \\ \hline
        \makecell[l]{Case $\mathcal{III}$:\\ $\beta=0$; $\gamma= 1$; $\mathring{A}^x\ne 0$; $\mathring{A}^t\ne 0$ or $\mathring{A}^z\ne 0$; $\mathring{A}^t\ne \mathring{A}^z$.} & \makecell[l]{$v_1=v_2$\\ \;\;\;\; =$v_3$} & \makecell[l]{$P_x\approx \frac{\mathring{A}^t-v\mathring{A}^z}{v\mathring{A}^x}P_b$\\ \qquad\;\; -$\frac{\mathring{A}^x(\mathring{A}^t-v\mathring{A}^z)}{2v}P_+$,\\ $P_l=2(\frac{1}{v^2}-1)P_b$.} & $P_+$, $P_x$, $P_+$, $P_b$, $P_l$. & 2
        \\ \hline
        \makecell[l]{Case $\mathcal{IV}$:\\ $\beta\ne \frac{1}{2}\gamma$; $\mathring{A}^x\ne 0$; $\mathring{A}^t=\mathring{A}^z$.\\ \;\; or\\ $\beta\ne 0$; $\gamma= 0$; $\mathring{A}^x\ne 0$ or $\mathring{A}^t\ne 0$ or $\mathring{A}^z\ne 0$.} & 1 & & $P_+$, $P_b$. & 2
        \\ \hline
        \makecell[l]{Case $\mathcal{V}$:\\ $\beta=\gamma=0$.\\ \;\; or\\ $\beta=\frac{1}{2}\gamma$ or $\mathring{A}^x=0$; $\mathring{A}^t=\mathring{A}^z$.} & 1 & & $P_+$. & 1
        \\ \hline
    \end{tabular}
    \end{footnotesize}
\caption{Polarization modes of GWs for the second group ($h_+, \Xi_x, \Theta, \Phi, \Sigma_x, \Omega, \Psi$) in case of $\Lambda_0=0$ and $\mu_0=0$ for general Einstein-vector theory.}
\label{Polarization3-table}
\end{table}

Specially, there are two special subcases for ``Case $\mathcal{I}$'' and ``Case $\mathcal{II}$'' in the speed $v_3$ that the polarization modes are different from the corresponding modes in TABLE~\ref{Polarization3-table}. They are
\begin{itemize}
    \item If $\beta\ne 0$ and $\gamma=\frac{1}{4}$, only $P_b$ mode is allowed in the speed $v_3$. And $v_3=1$.
    \item If $\beta=0$, no mode is allowed in the speed $v_3$.
\end{itemize}

\section{Analysis} \label{analysis}
In general Einstein-vector theory, the polarization modes of GWs depend on the chosen parameter spaces. The parameters include $\Lambda_0$, $\mu_0$, $\beta$, $\gamma$, and $\mathring{A}^{\mu}$.
Since we only consider a Minkowski background, so $\Lambda_0=0$. We also choose a coordinate system where GWs propagate along the ``+z'' direction and ``$\mathring{A}^y=0$''. Therefore, we only need to analyze six parameters. If $\mathring{A}^{\mu}=0$,
there are only two tensor modes. If $\mu_0=0$, the possible polarization modes of GWs depend on $\beta$, $\gamma$, $\mathring{A}^x$, $\mathring{A}^t$, and $\mathring{A}^z$. Based on the above discussions in the last section, 
we divide them into five cases (see TABLE~\ref{Polarization4-table}) according to the parameter spaces.\par
\begin{table}[h]
    \centering
    \begin{footnotesize}
    \begin{tabular}{|l|c|l|l|c|}\hline
        \textbf{Conditions} ($\Lambda_0=0$) & \textbf{Speed} & \textbf{Constraints} & \textbf{Modes} & \textbf{DoF} \\  \hline
        \multirow{3}*{\makecell[l]{Case 1:\\ $\mu_0=0$; $\gamma\ne 0,1$; $\mathring{A}^x\ne 0$; $\mathring{A}^t\ne 0$ or $\mathring{A}^z\ne 0$; $\mathring{A}^t\ne \mathring{A}^z$.}}
            & $v_1$ & \makecell[l]{$P_b\approx -\frac{\gamma \mathring{A}^x\mathring{A}^x}{2}P_+$,\\ $P_x=\frac{2(\mathring{A}^t-v\mathring{A}^z)}{v\mathring{A}^x}P_b$, \\$P_l=2(\frac{1}{v^2}-1)P_b$,\\ $P_y=\frac{(v^2-1)\mathring{A}^x}{v(v\mathring{A}^z-\mathring{A}^t)}P_{\times}$.} & \makecell[l]{$P_+$, $P_{\times}$,\\ $P_x$, $P_y$,\\ $P_b$, $P_l$.} & 2
            \\ \cline{2-5}
            & $v_2$ & \makecell[l]{$P_b=P_+$,\\ $P_x\approx\frac{\mathring{A}^t-v\mathring{A}^z}{v\mathring{A}^x}P_b$,\\ $P_l=2(\frac{1}{v^2}-1)P_b$,\\ $P_y=-\frac{v\mathring{A}^z-\mathring{A}^t}{v\mathring{A}^x}P_{\times}$.} & \makecell[l]{$P_+$, $P_{\times}$,\\ $P_x$, $P_y$,\\ $P_b$, $P_l$.} & 2
            \\ \cline{2-5}
            & $v_3$ & \makecell[l]{$P_+\approx -\frac{\gamma(4\gamma-1)\mathring{A}^x\mathring{A}^x}{2} P_b$,\\ $P_x=\frac{2(\mathring{A}^t-v\mathring{A}^z)}{v\mathring{A}^x}P_+$,\\ $P_l=2(\frac{1}{v^2}-1)P_b$.} & \makecell[l]{$P_+$,\\ $P_x$,\\ $P_b$, $P_l$.} & 1
        \\ \hline
        \multirow{6}*{\makecell[l]{Case 2:\\ $\mu_0=0$; $\gamma \ne 0$; $\mathring{A}^x=0$; $\mathring{A}^t\ne 0$ or $\mathring{A}^z\ne 0$; $\mathring{A}^t\ne \mathring{A}^z$.\\ \;\; or \\ $\mu_0=0$; $\beta\ne 0$; $\gamma=1$; $\mathring{A}^x\ne 0$; $\mathring{A}^t\ne 0$ or $\mathring{A}^z\ne 0$; $\mathring{A}^t\ne \mathring{A}^z$.}}
            & \multirow{2}*{$v_1$} & & \multirow{2}*{$P_+$, $P_{\times}$.} & \multirow{2}*{2}\\
            & & & &
            \\ \cline{2-5}
            & \multirow{2}*{$v_2$} & & \multirow{2}*{$P_x$, $P_y$.} & \multirow{2}*{2}\\
            & & & &
            \\ \cline{2-5}
            & \multirow{2}*{$v_3$} & \multirow{2}*{$P_l=2(\frac{1}{v^2}-1)P_b$.} & \multirow{2}*{$P_b$, $P_l$.} & \multirow{2}*{1}\\
            & & & &
        \\ \hline
        \makecell[l]{Case 3:\\ $\mu_0=0$; $\beta=0$; $\gamma= 1$; $\mathring{A}^x\ne 0$; $\mathring{A}^t\ne 0$ or $\mathring{A}^z\ne 0$; $\mathring{A}^t\ne \mathring{A}^z$.} & \makecell[l]{$v_1=v_2$\\ \;\;\;\; =$v_3$} & \makecell[l]{$P_x\approx \frac{\mathring{A}^t-v\mathring{A}^z}{v\mathring{A}^x}P_b$\\ \qquad\;\; -$\frac{\mathring{A}^x(\mathring{A}^t-v\mathring{A}^z)}{2v}P_+$,\\ $P_l=2(\frac{1}{v^2}-1)P_b$.} & \makecell[l]{$P_+$, $P_{\times}$,\\ $P_x$, $P_y$,\\ $P_b$, $P_l$.} & 4
        \\ \hline
        \makecell[l]{Case 4:\\ $\mu_0=0$; $\beta\ne \frac{1}{2}\gamma$; $\mathring{A}^x\ne 0$; $\mathring{A}^t=\mathring{A}^z$.\\ \;\;  or\\ $\mu_0=0$; $\beta\ne 0$; $\gamma=0$; $\mathring{A}^x\ne 0$ or $\mathring{A}^t\ne 0$ or $\mathring{A}^z\ne 0$.} & 1 & & \makecell[l]{$P_+$, $P_{\times}$,\\ $P_b$.} & 3
        \\ \hline
        \makecell[l]{Case 5:\\ $\mu_0=0$; $\beta=\gamma=0$.\\ \;\; or \;\; \\ $\mu_0=0$; $\beta=\frac{1}{2}\gamma$ or $\mathring{A}^x=0$; $\mathring{A}^t=\mathring{A}^z$.} & 1 & & $P_+$, $P_{\times}$. & 2
        \\ \hline
    \end{tabular}
    \end{footnotesize}
\caption{Polarization modes of GWs in general Einstein-vector theory.}
\label{Polarization4-table}
\end{table}
\indent Specially, for the speed $v_3$, there are two special subcases in both ``Case 1'' and ``Case 2'':
\begin{itemize}
    \item \textbf{Case 1.1:} $\beta\ne 0$ and $\gamma=\frac{1}{4}$ besides the conditions in case 1. Only the $P_b$ mode is allowed for the speed $v_3$, with $v_3=1$.
    \item \textbf{Case 1.2:} $\beta=0$ besides the conditions in case 1. No mode is allowed for the speed $v_3$.
    \item \textbf{Case 2.1:} $\beta\ne 0$ and $\gamma=\frac{1}{4}$ besides the conditions in case 2. Only the $P_b$ mode is allowed for the speed $v_3$, with $v_3=1$.
    \item \textbf{Case 2.2:} $\beta=0$ besides the conditions in case 2. No mode is allowed for the speed $v_3$.
\end{itemize}

\indent It is clear that all six modes of GWs in Eq.~\eqref{polarization-P} are possible in this theory, while the $P_l$ mode always appears in the form mixed with other modes. Additionally, many mixture polarization modes of GWs are possible in this gravity theory, including tensor-vector, tensor-scalar, tensor-vector-scalar, vector-scalar, and scalar-scalar modes. In each case, $P_+$ and $P_{\times}$ always exist, independently or mixed with other modes. We can also observe that the speed of the independent breathing mode is always the same as that of light (case 1.1, case 2.1, and case 4). In particular, when the speed of the tensorial GWs is strictly equal to that of light, besides the tensor modes, only the $P_b$ mode is allowed. 
By analyzing the speeds~\eqref{v1-2012}-\eqref{v3-2012}, we can find that the speed of GWs, which differs from that of light, primarily depends on $\gamma$ and $|\mathring{A}^t-\mathring{A}^z|$. The effect of $\mathring{A}^x$ on the speed of GWs is $\mathcal{O}(|\epsilon\mathring{A}^{\mu}\mathring{A}^{\nu}|^2)$, while the effect of $\mathring{A}^t$ and $\mathring{A}^z$ is $\mathcal{O}(|\epsilon\mathring{A}^{\mu}\mathring{A}^{\nu}|)$, where $\epsilon$ stands for $\gamma$, $\gamma^2$, or $\gamma(4\gamma-1)$. Therefore, the effect of $\mathring{A}^x$ on the speed of GWs is trivial compared to that of $\mathring{A}^t$ and $\mathring{A}^z$. 
One may ask why the scalar modes ($P_b$ and $P_l$) are only mixed with the plus mode ($P_+$), and not with the cross one ($P_{\times}$) ? This is because we have chosen a special coordinate with $\mathring{A}^y=0$. If we do not specialize $\mathring{A}^y$, the $P_{\times}$ mode will not be any more special in comparison to $P_+$.\par

We summarize the constraints between the amplitudes of polarization modes in the third column of TABLE~\ref{Polarization4-table}. According to the constraints, we have analyzed the order of magnitude of the amplitudes in Sec.~\ref{P-modes}. And we will combine these results with observations to restrict the parameter spaces in this theory (see Sec.~\ref{observation-850}). \par

\subsection{Comparison with other vector-tensor theories}
In Sec.~\ref{P-modes}, we have analyzed the polarization modes and the relations of the amplitudes in different parameter spaces and summarized the results in TABLE~\ref{Polarization4-table}. Here we will briefly introduce the polarization modes of GWs in Einstein-æther theory~\cite{Gong:2018cgj, Lin:2018ken}, generalized Proca theory~\cite{Dong:2023xyb}, and Bumblebee gravity theory~\cite{Liang:2022hxd}, and compare general Einstein-vector theory with them.\par

\textbf{Einstein-æther theory:} the action is given by Ref.~\cite{Jacobson:2004ts}. In this theory, the interaction of gravity is determined by the metric tensor $g_{\mu\nu}$ and the æther field $u^{\mu}$, while $u^{\mu}$ is a unit time-like vector field. When we consider the GWs propagating along the ``+z'' direction and assume that only the temporal component of the background vector field of $u^{\mu}$ is nonvanishing in the Minkowski background, there are six polarization modes of GWs in Einstein-æther theory, but only five independent degrees of freedom (See TABLE~\ref{Polarization5-table}). These modes are the $P_+$, $P_{\times}$, $P_x$, and $P_y$ modes, as well as the mixture of the $P_b$ and $P_l$ modes. Both Ref.~\cite{Gong:2018cgj} and Ref.~\cite{Lin:2018ken} have obtained the polarization modes of GWs in this theory. And the relations of the amplitudes of the polarization modes have been provided in Ref.~\cite{Lin:2018ken} : $\mathcal{O}$(vector) $\ll$ $\mathcal{O}$(scalar) $\ll$ $\mathcal{O}$(tensor).\par

\textbf{Generalized Proca theory:} this is also a vector-tensor theory~\cite{Heisenberg:2014rta}. The generalized Proca action depends on the metric tensor $g_{\mu\nu}$ and the vector field $A^{\mu}$. Consider a homogeneous and isotropic Minkowski background (meaning only the temporal component of the background vector field of $A^{\mu}$ is nonvanishing) and the GWs propagating along the ``+z'' direction. There are five independent modes of GWs: the $P_+$, $P_{\times}$, $P_x$, and $P_y$ modes, as well as the mixture of the $P_b$ and $P_l$ modes. This result has been provided by Ref.~\cite{Dong:2023xyb} (see TABLE~\ref{Polarization5-table}). For this theory, there are only $P_+$ and $P_{\times}$ in some parameter spaces. In other parameter spaces, besides the tensor modes, there are also two vector modes, or one scalar mode, or both of them. Additionally, in some parameter spaces, the mixture of the $P_b$ and $P_l$ modes will turn into the $P_b$ mode.\par
 
\textbf{Bumblebee gravity theory:} this theory introduces a dynamical vector field coupled to gravity, and the action is given by Ref.~\cite{Kostelecky:2003fs}. Also consider the GWs propagating along the ``+z'' direction and an arbitrary background vector field $b^{\mu}$ in the Minkowski background. The polarization modes of GWs in this theory have been studied in Ref.~\cite{Liang:2022hxd}. The polarization modes of GWs depend on the parameter spaces, which have been divided into three cases in Ref.~\cite{Liang:2022hxd} (see TABLE~\ref{Polarization5-table}). In one case, there are two mixture modes: one mixture of the $P_+$, $P_x$, $P_y$, $P_b$, and $P_l$ modes, and another mixture of the $P_{\times}$, $P_x$, $P_y$, $P_b$, and $P_l$ modes. In another case, there are five independent modes: the $P_+$, $P_{\times}$, $P_x$, and $P_y$ modes, as well as the mixture of the $P_b$ and $P_l$ modes. And when $b^t= b^z$, there are only the $P_+$ and $P_{\times}$ modes with the same speed as light. Specifically,  the cases where vector modes or scalar modes are not suppressed are ignored in Ref.~\cite{Liang:2022hxd}. 
\begin{table}[h]
    \centering
    \begin{footnotesize}
        \begin{tabular}{|l|l|l|c|}
            \hline
            \textbf{Gravity theory} & \textbf{Cases} & \textbf{Modes} & \textbf{DoF}\\ 
            \hline
            Einstein-æther theory & $\mathring{u}^{\mu}=(1,0,0,0)$. & $P_+$, $P_{\times}$, $P_x$, $P_y$, $P_b$, $P_l$. & 5 \\
            \hline
            \multirow{3}*{Generalized Proca theory}
                & \textbf{i:} $\mathring{A}^{\mu}=(A,0,0,0)$, $\mathring{G}_4-\mathring{G}_{4,X}A^2\ne 0$. & $P_+$, $P_{\times}$. & 2 
                \\ \cline{2-4}
                & \makecell[l]{\textbf{ii:} $\mathring{A}^{\mu}=(A,0,0,0)$, $\mathring{G}_{4,X}=0$, $1-2c_2\mathring{G}_{4,X}\ne 0$,\\ \quad\; $(1-2c_2\mathring{G}_{4,X})(\mathring{G}_4-\mathring{G}_{4,X}A^2)+\mathring{G}^2_{4,X}A^2\ne 0$.} & $P_x$, $P_y$. & 2 
                \\ \cline{2-4}
                & \makecell[l]{\textbf{iii:} $\mathring{A}^{\mu}=(A,0,0,0)$, $1-2c_2\mathring{G}_{4,X}\ne 0$,\\ \quad\;\; $\mathring{G}_{3,X}\ne 0$, $\mathcal{C}\ne 0$, $\mathcal{C}_2\ne 0$.} & $P_b$, $P_l$. & 1
            \\ \hline
            \multirow{3}*{Bumblebee gravity theory}
                & \textbf{i:} $b_x^2+b_y^2\ne 0$, $b^t\ne b^z$. & $P_{\times}$, $P_+$, $P_x$, $P_y$, $P_b$, $P_l$. & 2 
                \\ \cline{2-4}
                & \textbf{ii:} $b^x=b^y=0$, $b^t\ne b^z$. & $P_{\times}$, $P_+$, $P_x$, $P_y$, $P_b$, $P_l$. & 5 
                \\ \cline{2-4}
                & \textbf{iii:} $b^t= b^z$. & $P_+$, $P_{\times}$. & 2
            \\ \hline
        \end{tabular}
    \end{footnotesize}
\caption{Polarization modes of Einstein-æther theory~\cite{Gong:2018cgj, Lin:2018ken}, generalized Proca theory~\cite{Dong:2023xyb}, and Bumblebee gravity theory~\cite{Liang:2022hxd}. The third column lists the possible polarization modes. The last column gives the degree of freedom of the polarization modes. (Notion: these cases in the second column contain different meanings for different theories. For Einstein-æther theory and Bumblebee theory, these cases represent different parameter spaces. However, for generalized Proca theory, case i, case ii, and case iii represent the conditions for the presence of tensor modes, vector modes, and scalar modes, respectively.)}
\label{Polarization5-table}
\end{table}

Combining the polarization modes of Einstein-æther theory, Bumblebee gravity theory, and general Einstein-vector theory (as shown in TABLEs~\ref{Polarization4-table} and~\ref{Polarization5-table}), we can observe that these vector-tensor theories have similar polarization modes. If only the temporal component of the background vector field is nonzero, all these theories allow five independent modes (the $P_+$, $P_{\times}$, $P_x$, and $P_y$ modes, as well as the mixture of the $P_b$ and $P_l$ modes), when we do not consider the special values of the other parameters. And if the temporal component of the background vector field is also zero, they all allow only $P_+$ and $P_{\times}$ modes, which is the same as that of general relativity.\par

\textbf{Comparison with generalized Proca theory:} if only the temporal component of the background vector field is nonzero, the properties of polarizations are quite similar between generalized Proca theory and general Einstein-vector theory. Both theories allow four types of polarization in different parameter spaces: \ding{172} two tensor modes, two vector modes, and one scalar mode; \ding{173} two tensor modes and one scalar mode; \ding{174} two tensor modes and two vector modes; \ding{175} two tensor modes.\par

\textbf{Comparison with Bumblebee gravity theory:} Bumblebee gravity theory and general Einstein-vector theory were studied with an arbitrary background vector field in a Minkowski background. By combining TABLEs~\ref{Polarization4-table} and~\ref{Polarization5-table}, we can find that ``Case 1'' in TABLE~\ref{Polarization4-table} corresponds to ``Case i'' in Bumblebee gravity theory, as both allow the mixture of the tensor, vector, and scalar modes. ``Case 2'' corresponds to ``Case ii'', allowing the two tensor modes, two vector modes, and one scalar mode with different speeds. ``Case 5'' corresponds to ``Case iii'', permitting only the two tensor modes with the same speed as light. In general Einstein-vector theory, there are two additional parameters, $\beta$ and $\gamma$, which significantly impact the polarization modes. In Ref.~\cite{Liang:2022hxd}, the cases that vector modes or scalar modes are not suppressed were ignored in Bumblebee gravity theory. Therefore, within the context of general Einstein-vector theory, we now assume that $A^y$ is not distinct from $A^x$. We consider scenarios where neither $\beta$ nor $\gamma$ takes on special values and disregard all instances where vector modes or scalar modes are not suppressed. Then ``Case 1'' allows for only two independent mixture modes, similar to those in Bumblebee gravity theory. The same applies to both ``Case 2'' and ``Case 5''. Switching the $t$ and $z$ components of the background vector field has no effect on the polarization results in these two theories.\par

Based on the above analysis, we can observe that there are numerous similarities in the polarizations of these four vector-tensor gravity theories. When considered under the same conditions, the similarities are quite striking.

\subsection{Observation and analysis} \label{observation-850}
On August 17, 2017, a binary neutron star coalescence candidate (GW170817)~\cite{LIGOScientific:2017vwq} was observed through GWs by Advanced LIGO and Virgo. About 1.7 seconds later, the Fermi Gamma-ray Burst Monitor independently detected a gamma-ray burst (GRB170817A)~\cite{Goldstein:2017mmi}. This marks the first multi-messenger observation of a binary neutron star merger. The observations place a tight constraint on the speed ($c_T$) of the tensorial GWs~\cite{abbott2017gravitational, abbott2019tests}: 
\begin{eqnarray}
    -3 \times 10^{-15} \le c_T-1 \le 7 \times 10^{-16}.\label{ct-1738}
\end{eqnarray}
According to TABLE~\ref{Polarization4-table}, if the constraint~\eqref{ct-1738} is assumed to be valid for all tensor modes, then we can obtain some constraints for parameters in general Einstein-vector theory:
\begin{eqnarray}
    \textbf{Case 1:}&& -6\times 10^{-15}\le \gamma(\mathring{A}^t-\mathring{A}^z)^2\le 1.4\times 10^{-15},\;\; -6\times 10^{-15}\le \gamma^2(\mathring{A}^t-\mathring{A}^z)^2\le 1.4\times 10^{-15},\nonumber\\ 
    && -1.8\times 10^{-14}\le \gamma(4\gamma-1)(\mathring{A}^t-\mathring{A}^z)^2\le 4.2\times 10^{-15};\nonumber\\
    \textbf{Case 2:}&& -6\times 10^{-15}\le \gamma(\mathring{A}^t-\mathring{A}^z)^2\le 1.4\times 10^{-15};\nonumber\\
    \textbf{Case 3:}&& -6\times 10^{-15}\le (\mathring{A}^t-\mathring{A}^z)^2\le 1.4\times 10^{-15};\nonumber\\
    \textbf{Case 4:}&& \text{no constraint};\nonumber\\
    \textbf{Case 5:}&& \text{no constraint}.\nonumber
\end{eqnarray}
\indent Specifically, when the speed of the tensorial GWs is strictly equal to that of light, in addition to tensor modes, only $P_b$ is allowed. And only the parameter spaces ``\textbf{Case 4 $\bigcup$ Case 5}'' are viable.
\par
Last year, PTAs announced compelling evidence for a stochastic signal in their latest data sets~\cite{NANOGrav:2023gor,EPTA:2023fyk,Reardon:2023gzh,Xu:2023wog}. This implies PTA may open new windows to explore the nano-hertz frequency range of the spectrum. Reference~\cite{Chen:2023uiz} searched for an isotropic nontensorial gravitational-wave background, which is allowed by general metric theories of gravity, in the North American Nanoherz Observatory for GWs 15-year data set. The study reveals that the signal of the transverse scalar mode of GWs very possibly exists according to the data set, and the amplitude of this transverse scalar mode is equal in magnitude to that of the tensor modes. There are also other relevant studies that investigate the scalar modes based on the data of PTAs~\cite{NANOGrav:2023ygs}. \par

According to the analysis in Sec.~\ref{P-modes} and combining TABLE~\ref{Polarization4-table}, we find that there exist potential solutions within the parameter spaces ensure either the absence of other modes or the suppression of other modes relative to the tensor modes. However, based on the findings in Ref.~\cite{Chen:2023uiz}, we anticipate that GWs will permit the amplitude of the $P_b$ mode to be equivalent in magnitude to that of the tensor modes, while the $P_l$, $P_x$, and $P_y$ modes are expected to be either suppressed or disallowed. \par

Based on the expectation that the amplitude of the $P_b$ mode is equivalent in magnitude to that of the tensor modes, and in accordance with the aforementioned analysis for general Einstein-vector theory, we can find
\begin{itemize}
    \item If we assume that the $P_l$, $P_x$, and $P_y$ modes are suppressed or not permitted, then the parameter spaces ``\textbf{Case 1.1 $\bigcup$ Case 2 $\bigcup$ Case 4}'' remain viable.
    \item If we just assume that the $P_l$, $P_x$, and $P_y$ modes are not permitted, then only the parameter spaces ``\textbf{Case 2.1 $\bigcup$ Case 4}'' remain viable.
    \item If assume that the $P_l$, $P_x$, and $P_y$ modes are not permitted and the speed of the tensorial GWs is strictly equal to that of light, then only the parameter spaces ``\textbf{Case 4}'' remain viable.
\end{itemize}

\indent We summarize the results in TABLE~\ref{constraint-1050}.
\begin{table}[h]
    \centering
        \begin{tabular}{|c|c|c|c|c|}
            \hline
            \textbf{Conditions} & $v_{\text{tensor}}=1$ & \textbf{Assumption \ding{172}} & \textbf{Assumption \ding{173}} & \textbf{Assumption \ding{174}} \\  \hline
            \textbf{Case 1} & -       & Case 1.1     & -        & -  \\ 
            \hline
            \textbf{Case 2} & -       & $\beta\ne 0$ & Case 2.1 & - \\
            \hline
            \textbf{Case 3} & -       & -            & -        & - \\
            \hline
            \textbf{Case 4} & $\surd$ & Case 4       & Case 4   & Case 4 \\ 
            \hline
            \textbf{Case 5} & $\surd$ & -            & -        & - \\
            \hline
        \end{tabular}
\caption{The constraints on parameter spaces for general Einstein-vector theory under three assumptions: \ding{172} The $P_b$ mode remains unsuppressed, while the $P_l$, $P_x$, and $P_y$ modes are either suppressed or disallowed; \ding{173} The $P_b$ mode remains unsuppressed, while the $P_l$, $P_x$, and $P_y$ modes are disallowed; \ding{174} The $P_b$ mode remains unsuppressed, while the $P_l$, $P_x$, and $P_y$ modes are disallowed and the speed of the tensorial GWs is strictly equal to that of light. The ``$\surd$'' in the second column means the speed of the tensorial GWs is strictly equal to that of light in the corresponding parameter spaces. The conditions in the last three columns are the cases that satisfy the corresponding assumptions. And the ``-" in this table means that the corresponding cases are not viable.}
\label{constraint-1050}
\end{table}

We cannot overlook the importance of the mathematical consistency of modified gravity theories. The minimal consistency requirement for a classical theory is that the initial value problem must be locally well-posed. This means that there should exit a unique solution of the equations of motion for given suitable initial data, and this solution should depend continuously on the data~\cite{Papallo:2017qvl}. For the linearized equations of motion, a well-posed initial value problem requires that the equations should be hyperbolic, which means they possess the character of a wave equation. For the weak and strong hyperbolicities, see Refs.~\cite{Papallo:2017qvl, kreiss1989initial, Sarbach:2012pr}.
\par
For general Einstein-vector theory considered in this paper, there are two sets of linearized equations of motion~\eqref{linear1-A}-\eqref{perturbationEq-g}. Firstly, if $\Lambda_0=0$ and $\mathring{A}^{\mu}=0$, the linearized equations~\eqref{linear1-A} and~\eqref{linear1-g} imply that there is no interaction between the vector field and the metric field under linear approximations in the Minkowski background. So we can discuss their hyperbolicity respectively. For the linearized equation of motion of the metric~\eqref{linear1-g}, It is the linearized vacuum Einstein equation. It is well-known to be strongly hyperbolic in harmonic gauge which means a well-posed initial value problem. For the linearized equation of motion of the vector field~\eqref{linear1-A}, the coefficient matrix of the term $\partial_t\partial_t a_{\mu}$ is irreversible. This is because we do not impose the gauge condition. The gauge freedom implies that the equation will not be hyperbolic. However, because of the mass term, the Lorenz gauge condition does not work in this case. Secondly, if $\Lambda_0=0$ and $\mu_0=0$, the linearized equations of motion are Eqs.~\eqref{perturbationEq-a} and~\eqref{perturbationEq-g}. The result of the observations requires that the coupling constants $\gamma$ and $\beta$ should be very small. Considering $\gamma\rightarrow 0$ and $\beta\rightarrow 0$, then it becomes the standard Einstein-Maxwell theory. And it is strongly hyperbolic when one imposes the modified-harmonic-gauge condition~\cite{Davies:2021frz}.
\par
Since the anisotropy of space and the broken of $U(1)$ gauge symmetry in general Einstein-vector theory, its hyperbolicity and the corresponding gauge condition may exhibit many interesting differences from other gravity theories. It is meaningful to study the local well-posedness of the initial value problem for this gravity theory, especially for the case of Eq.~\eqref{linear1-A} and the case of Eqs.~\eqref{perturbationEq-a} and~\eqref{perturbationEq-g} when $\gamma,\beta \ne 0$. The local well-posedness of the initial value problem will imply whether it is suitable to solve the equations of motion numerically on a computer.

\section{Conclusion} \label{conclusion}
An increasing number of unresolved issues in general relativity have led to the rise of modified gravity theories as a critical route towards uncovering the ultimate theory of gravity.
In recent years, the tensor modes of GWs have been detected~\cite{LIGOScientific:2017ycc}, and the data of PTAs indicates that the existence of the breathing mode is probable~\cite{Chen:2023uiz}. This suggests that polarization modes of GWs could potentially be used to test modified gravity theories in the foreseeable future. 
General Einstein-vector theory is an intriguing candidate for explaining dark energy, dark matter, and the inflationary universe~\cite{Geng:2015kvs}. In this theory, the vector field couples bilinearly to 
curvature. This theory has many advantages, including ghost-free, the emergence of $U(1)$ gauge symmetry in certain weak-field limits, and two-derivative field equations under linear approximations.\par

In this paper, we investigated the polarization modes of GWs within general Einstein-vector theory, incorporating an arbitrary background vector field under a Minkowski background. Firstly, we obtained the field equations from the action. 
Then using linear approximations, we derived the background equations and the linear perturbation equations. After that, we analyzed the polarizations of GWs in general Einstein-vector theory and their amplitude relations using Bardeen framework. Next, we compared the polarization modes of this theory with those of Einstein-æther theory, generalized Proca theory, and Bumblebee gravity theory. Finally, we briefly analyzed the speed of GWs and restricted the parameter spaces of general Einstein-vector theory by the gravitational-wave event GW170817 with its electromagnetic counterpart GRB170817A and the observations of PTAs.\par

Considering an arbitrary background vector field, the space is anisotropic and the loss of space-time symmetry will lead to some interesting phenomena. We found that the polarization modes of GWs depend on the values of $\mu_0$, $\beta$, $\gamma$, and $\mathring{A}^{\mu}$. 
All six possible independent modes of the metric theory have appeared, while the $P_l$ mode always appears in a manner mixed with other modes. And many mixture polarization modes are possible in this gravity theory, including tensor-vector, tensor-scalar, tensor-vector-scalar, vector-scalar, and scalar-scalar modes. Regardless of the values of these parameters, the two tensor modes ($P_+$ and $P_{\times}$) are always allowed either independent of or mixed with other modes. We found that the speed of the independent breathing mode is always the same as
that of light. And if the speed of the tensorial GWs is strictly equal to that of light, besides tensor
modes, there is only the $P_b$ mode allowed. For the speed of GWs that is different from that of light, it mainly depends on $\gamma$ and $|\mathring{A}^t-\mathring{A}^z|$ (see Eqs.~\eqref{v1-2012}-\eqref{v3-2012}).
Considering that the speed of tensorial GWs is the same as that of light based on the observation of GW170817 and GRB170817A, the parameter spaces ``Case 4 $\bigcup$ Case 5'' are viable. Considering that the amplitude of the $P_b$ mode is the same magnitude as that of the tensor modes and no other modes based on the data of PTAs, the parameter spaces ``Case 2.1 $\bigcup$ Case 4'' are viable. Then if we consider the two constraints, only the parameter spaces ``Case 4'' are viable. More detailed results can be found in TABLE~\ref{constraint-1050}.\par

Our study provides a theoretical understanding for vector-tensor gravity in the case of an arbitrary background vector field. The loss of space-time symmetry caused by the vector field results in abundant content of GWs. The comparisons of vector-tensor theories can provide a deeper understanding of these theories. The polarization modes of each vector-tensor theory and the restrictions of the gravitational-wave event GW170817 with its electromagnetic counterpart GRB170817A and the observations of PTAs make us expect to the confirmation of the breathing mode.
With the continuous detection of ground-based gravitational wave detections (LIGO, Virgo, KAGRA, PTAs, FAST, and others) and the progress of the Lisa, Taiji, and TianQin projects~\cite{LISA:2017pwj,10.1093.ptep.ptaa083,TianQin:2015yph}, this gravity theory is expected to be tested and 
its parameter spaces are expected to be selected.

\section*{Acknowledgments}
We would like to thank Chun-Chun Zhu, Shan-Ping Wu, and Yu-Peng Zhang for useful discussions. 
This work was supported by 
the National Natural Science Foundation of China (Grants No.~12475056, No.~123B2074 and No.~12247101), 
the 111 Project (Grant No.~B20063), 
the Major Science and Technology Projects of Gansu Province, 
and Lanzhou City's scientific research funding subsidy to Lanzhou University.\par

\appendix
\section{Field equations consisting of gauge invariants} \label{Appendix1}
Substituting Eq.~\eqref{decomposition-ha} into the perturbation equation~\eqref{perturbationEq-a} and applying the method of gauge invariants, we can derive the perturbation equation of the vector field consisting of gauge invariants:
\begin{eqnarray}
    f^t&\equiv&(1-\gamma)\mathring{A}^{j}\nabla^{2}\Xi_{j}+\big[2(2\beta-\gamma)\mathring{A}^{t}\nabla^{2}-(1+2\gamma)\mathring{A}^{j}\partial_{j}\partial_{t}
    -6\beta\mathring{A}^t\partial_t\partial_t\big]\Theta+2(2\beta-1)\mathring{A}^{t}\nabla^{2}\Phi\nonumber\\
    && -\nabla^{2}\Omega-\partial_{t}\nabla^{2}\Psi =0 ,\\
    f^i&\equiv&(\gamma-1)\delta^{ik}\mathring{A}^{j}\overline{\square}h^{TT}_{kj}-\overline{\square}\Sigma^{i}+\mathring{A}^{t}\big(\gamma \overline{\nabla}^2-\overline{\square}\big)\Xi^{i}
    +(\gamma-1)\delta^{ik}\mathring{A}^{j}\partial_{k}\partial_{t}\Xi_{j} +\gamma\mathring{A}^{j}\partial_{j}\partial_{t}\Xi^{i}+\delta^{ik}\partial_{k}\partial_{t}\Omega
    +\delta^{ik}\partial_{k}\partial_{t}\partial_{t}\Psi\nonumber\\
    && +2(2\beta-\gamma)\mathring{A}^{i}\nabla^{2}\Phi+ 2\gamma\delta^{ik}\mathring{A}^{j}\partial_{k}\partial_{j}\Phi
    +2\delta^{ik}\mathring{A}^{t}\partial_{k}\partial_{t}\Phi +2\gamma\delta^{ik}\mathring{A}^t \partial_k\partial_t\Theta +(\gamma+1)\delta^{ik}\mathring{A}^j\partial_k\partial_j\Theta\nonumber\\
    && -\mathring{A}^i\big[(2\beta-\gamma)\partial_t\partial_t+ (1+\gamma-4\beta)\overline{\square}\big]\Theta =0.
\end{eqnarray}
Using the same method, we can derive the perturbation equation of the metric field consisting of gauge invariants:
\begin{eqnarray}
    \overset{\text{{\fontsize{3.5pt}{\baselineskip}\selectfont \textbf{(1)}}}}{G}_{\mu\nu}-\beta \overset{\text{{\fontsize{3.5pt}{\baselineskip}\selectfont \textbf{(1)}}}}{Y}_{\mu\nu}-\gamma \overset{\text{{\fontsize{3.5pt}{\baselineskip}\selectfont \textbf{(1)}}}}{Z}_{\mu\nu}=0.
\end{eqnarray}
The specific forms of $\overset{\text{{\fontsize{3.5pt}{\baselineskip}\selectfont \textbf{(1)}}}}{G}_{\mu\nu}$, $\overset{\text{{\fontsize{3.5pt}{\baselineskip}\selectfont \textbf{(1)}}}}{Y}_{\mu\nu}$, and $\overset{\text{{\fontsize{3.5pt}{\baselineskip}\selectfont \textbf{(1)}}}}{Z}_{\mu\nu}$ are\\
$\overset{\text{{\fontsize{3.5pt}{\baselineskip}\selectfont \textbf{(1)}}}}{G}_{\mu\nu}$ :
\begin{eqnarray}
    \overset{\text{{\fontsize{3.5pt}{\baselineskip}\selectfont \textbf{(1)}}}}{G}_{tt}&=&-\overline{\nabla}^2\Theta, \\
    \overset{\text{{\fontsize{3.5pt}{\baselineskip}\selectfont \textbf{(1)}}}}{G}_{ti}&=&-\frac{1}{2}\overline{\nabla}^2\Xi_i-\Theta_{ti},\\
    \overset{\text{{\fontsize{3.5pt}{\baselineskip}\selectfont \textbf{(1)}}}}{G}_{ij}&=&\frac{1}{2}\Big( 2\delta_{ij}\nabla^{2}\Phi-2\partial_{i}\partial_{j}\Phi-\partial_{i}\partial_{t}\Xi_{j}-\partial_{j}\partial_{t}\Xi_{i}
        +\partial_{t}\partial_{t}h^{TT}_{ij} -\nabla^{2}h^{TT}_{ij}-2\delta_{ij}\partial_{t}\partial_{t}\Theta-\partial_{i}\partial_{j}\Theta+\delta_{ij}\nabla^{2}\Theta \Big).
\end{eqnarray}
$\overset{\text{{\fontsize{3.5pt}{\baselineskip}\selectfont \textbf{(1)}}}}{Y}_{\mu\nu}$ :
\begin{eqnarray}
    \overset{\text{{\fontsize{3.5pt}{\baselineskip}\selectfont \textbf{(1)}}}}{Y}_{tt}&=&\mathring{A}^{i}\mathring{A}^{j}\nabla^{2}h^{TT}_{ij} +2\mathring{A}^{i}\mathring{A}^{t}\overline{\nabla}^2\Xi_{i}+2\mathring{A}^{i}\nabla^{2}\Sigma_{i}
        +2\mathring{A}^{i}\partial_{i}\nabla^{2}\Psi-2\mathring{A}^{t}\nabla^{2}\Omega+A^{t}A^{t}(\nabla^{2}-3\partial_{t}\partial_{t})\Theta \nonumber\\
        && +2A^{i}A_{i}\nabla^{2}\Theta,\\
        \overset{\text{{\fontsize{3.5pt}{\baselineskip}\selectfont \textbf{(1)}}}}{Y}_{ti}&=&\mathring{A}^{j}\mathring{A}^{k}\partial_{i}\partial_{t}h^{TT}_{jk} +2\mathring{A}^{j}\partial_{i}\partial_{t}\Sigma_{j} +\frac{1}{2}\mathring{A}^{\mu}\mathring{A}_{\mu}\nabla^{2}\Xi_{i}
        +2\mathring{A}^{j}\mathring{A}^{t}\partial_{i}\partial_t\Xi_{j}-2\mathring{A}_{i}\mathring{A}^{t}\nabla^{2}\Phi -2\mathring{A}^{t}\mathring{A}^{t}\partial_{i}\partial_{t}\Phi \nonumber\\
        &&  -2A^{t}\partial_{i}\partial_{t}\Omega +2A^{j}\partial_{j}\partial_{i}\partial_{t}\Psi-2\mathring{A}_{i}\mathring{A}^{t}\overline{\square}\Theta
        +\mathring{A}_{i}\mathring{A}^{t}\partial_{t}\partial_{t}\Theta +2\mathring{A}^{j}\mathring{A}_{j}\partial_{i}\partial_{t}\Theta -\mathring{A}^{t}\mathring{A}^{t}\partial_{i}\partial_{t}\Theta, \\
        \overset{\text{{\fontsize{3.5pt}{\baselineskip}\selectfont \textbf{(1)}}}}{Y}_{ij}&=&-\mathring{A}^{\mu}\mathring{A}_{\mu}\overset{\text{{\fontsize{3.5pt}{\baselineskip}\selectfont \textbf{(1)}}}}{G}_{ij} +\mathring{A}^{k}\mathring{A}^{l}\big(\partial_{i}\partial_{j} -\delta_{ij}\overline{\square}\big)h^{TT}_{kl} +2A^{k}\big(\partial_{i}\partial_{j}
        -\delta_{ij}\overline{\square}\big)\Sigma_{k}+2\mathring{A}^{k}\mathring{A}^{t}\big(\partial_{i}\partial_{j} -\delta_{ij}\overline{\square}\big)\Xi_{k} \nonumber\\
        && -2\mathring{A}^{t}\big(\partial_{i}\partial_{j} -\delta_{ij}\overline{\square}\big)\Omega+2\mathring{A}^{k}\partial_{k}\big(\partial_{i}\partial_{j} -\delta_{ij}\overline{\square}\big)\Psi
        +2\mathring{A}_{i}\mathring{A}_{j}\nabla^{2}\Phi -2\mathring{A}^{t}\mathring{A}^{t}\big(\partial_{i}\partial_{j} -\delta_{ij}\overline{\square}\big)\Phi \nonumber\\
        && +\mathring{A}^{k}\mathring{A}_{k}\big(\partial_{i}\partial_{j} -\delta_{ij}\overline{\square}\big)\Theta +2\mathring{A}_{i}\mathring{A}_{j}\overline{\square}\Theta -\mathring{A}_i\mathring{A}_j\partial_t\partial_t\Theta.
\end{eqnarray}
$\overset{\text{{\fontsize{3.5pt}{\baselineskip}\selectfont \textbf{(1)}}}}{Z}_{\mu\nu}$ :
\begin{eqnarray}
    \overset{\text{{\fontsize{3.5pt}{\baselineskip}\selectfont \textbf{(1)}}}}{Z}_{tt}&=&-\frac{1}{2}\mathring{A}^{i}\mathring{A}^{j}\nabla^{2}h^{TT}_{ij} -\mathring{A}^{i}\nabla^{2}\Sigma_{i} -\mathring{A}^{i}\mathring{A}^{t}\nabla^{2}\Xi_{i}
        -\frac{1}{2}\mathring{A}^{t}\mathring{A}^{t}\overline{\square}\Theta -\mathring{A}^{i}\mathring{A}_{i}\nabla^{2}\Theta-\frac{1}{2}\mathring{A}^{t}\mathring{A}^{t}\partial_{t}\partial_{t}\Theta
        +\frac{3}{2}\mathring{A}^{i}\mathring{A}^{j}\partial_{i}\partial_{j}\Theta,
        \\
        \overset{\text{{\fontsize{3.5pt}{\baselineskip}\selectfont \textbf{(1)}}}}{Z}_{ti}&=&-\frac{1}{2}\mathring{A}^j\mathring{A}^k\partial_i\partial_t h^{TT}_{jk} + \frac{1}{2}\mathring{A}^t\mathring{A}^j\partial_j\partial_t\Xi_i +\frac{1}{2}\mathring{A}^{j}\mathring{A}^k\partial_j\partial_k\Xi_i
        -\frac{1}{4}\mathring{A}^{\mu}\mathring{A}_{\mu}\overline{\nabla}^2\Xi_i-\mathring{A}^t\mathring{A}^j\partial_i\partial_t\Xi_j +\frac{1}{2}\mathring{A}^{t}\nabla^{2}\Sigma_{i} \nonumber \\
        && +\frac{1}{2}\mathring{A}^{j}\partial_{j}\partial_{t}\Sigma_{i} -\mathring{A}^{j}\partial_{i}\partial_{t}\Sigma_{j} +\frac{1}{2}\mathring{A}_{i}\nabla^{2}\Omega-\frac{1}{2}\mathring{A}^{j}\partial_{j}\partial_{i}\Omega
        -\frac{1}{2}\mathring{A}^{j}\partial_{j}\partial_{i}\partial_{t}\Psi +\frac{1}{2}\mathring{A}_{i}\nabla^{2}\partial_{t}\Psi +\mathring{A}^{t}\mathring{A}_{i}\nabla^{2}\Phi \nonumber\\
        && -\mathring{A}^{t}\mathring{A}^{j}\partial_{i}\partial_{j}\Phi-\frac{1}{2}\mathring{A}^{t}\mathring{A}^{t}\partial_{i}\partial_{t}\Theta -\mathring{A}^{j}\mathring{A}_j\partial_{i}\partial_{t}\Theta-\mathring{A}^{j}\mathring{A}^t\partial_{i}\partial_{j}\Theta +\mathring{A}_{i}\mathring{A}^{t}\overline{\nabla}^2\Theta +\frac{3}{2}\mathring{A}^j\mathring{A}_i\partial_{j}\partial_{t}\Theta,
        \\
        \overset{\text{{\fontsize{3.5pt}{\baselineskip}\selectfont \textbf{(1)}}}}{Z}_{ij}&=&\frac{1}{2}\eta_{ij}\mathring{A}^k\mathring{A}^l\overline{\square}h^{TT}_{kl} -\frac{1}{4}\mathring{A}^{\mu}\mathring{A}_{\mu}\overline{\square} h^{TT}_{ij}
        +\frac{1}{2}\mathring{A}^{\alpha}\mathring{A}^{\rho}\partial_{\alpha}\partial_{\rho}h^{TT}_{ij} -\frac{1}{2}\mathring{A}^{k}\mathring{A}^{l}\partial_{i}\partial_{j}h^{TT}_{kl}
        -\frac{1}{4}\eta_{ij}\mathring{A}^{\alpha}\mathring{A}_{\alpha}\overline{\square}\Theta \nonumber\\
        && +\frac{1}{2}\eta_{ij}\mathring{A}^{\alpha}\mathring{A}_{\alpha}\nabla^{2}\Theta+\frac{1}{2}\eta_{ij}\mathring{A}^{k}\mathring{A}_{k}\overline{\square}\Theta -\frac{3}{4}\eta_{ij}\mathring{A}^{k}\mathring{A}_{k}\partial_{t}\partial_{t}\Theta
        -\frac{1}{4}\eta_{ij}\mathring{A}^{\alpha}\mathring{A}^{\rho}\partial_{\alpha}\partial_{\rho}\Theta -\frac{3}{2}\eta_{ij}\mathring{A}^k\mathring{A}^t\partial_k\partial_t\Theta \nonumber\\
        && -\frac{3}{4}\eta_{ij}\mathring{A}^k\mathring{A}^l\partial_k\partial_l\Theta-\mathring{A}_i\mathring{A}_j\overline{\square}\Theta+\frac{1}{2}\mathring{A}_i\mathring{A}_j\partial_t\partial_t\Theta -\frac{1}{4}\mathring{A}^{\alpha}\mathring{A}_{\alpha}\partial_i\partial_j\Theta -\frac{1}{2}\mathring{A}^k\mathring{A}_k\partial_i\partial_j\Theta
        +2\mathring{A}^{\alpha}\mathring{A}_{(i}\partial_{j)}\partial_{\alpha}\Theta \nonumber\\
        && -\eta_{ij}\mathring{A}^t\mathring{A}^t\overline{\nabla}^2\Phi+\frac{1}{2}\eta_{ij}\mathring{A}^{\alpha}\mathring{A}_{\alpha}\overline{\nabla}^2\Phi-\eta_{ij}\mathring{A}^k\mathring{A}^l\partial_k\partial_l\Phi -2\eta_{ij}\mathring{A}^k\mathring{A}^t\partial_k\partial_t\Phi-\mathring{A}_i\mathring{A}_j\overline{\nabla}^2\Phi +\mathring{A}^t\mathring{A}^t\partial_i\partial_j\Phi \nonumber\\
        && -\frac{1}{2}\mathring{A}^{\alpha}\mathring{A}_{\alpha}\partial_i\partial_j\Phi+2\mathring{A}^{\alpha}\mathring{A}_{(i}\partial_{j)}\partial_{\alpha}\Phi-\eta_{ij}\mathring{A}^{k}\partial_{k}\partial_{t}\Omega +\mathring{A}^{t}\partial_{i}\partial_{j}\Omega +\mathring{A}_{(i}\partial_{j)}\partial_{t}\Omega -\eta_{ij}\mathring{A}^{t}\nabla^{2}\Omega \nonumber\\
        && -\eta_{ij}\mathring{A}^k\partial_k\partial_t\partial_t\Psi-\eta_{ij}\mathring{A}^t\overline{\nabla}^2\partial_t\Psi+\mathring{A}^t\partial_t\partial_i\partial_j\Psi+\mathring{A}_{(i}\partial_{j)}\partial_{t}\partial_{t}\Psi +\eta_{ij}\mathring{A}^{k}\overline{\square}\Sigma_{k} -\mathring{A}_{(i}\overline{\square}\Sigma_{j)} \nonumber\\
        && -\mathring{A}^{k}\partial_i\partial_{j}\Sigma_{k} +\mathring{A}^{t}\partial_{t}\partial_{(i}\Sigma_{j)}+\mathring{A}^{k}\partial_{k}\partial_{(i}\Sigma_{j)}+\eta_{ij}\mathring{A}^k\mathring{A}^t\overline{\square}\Xi_k-\mathring{A}^{t}\mathring{A}_{(i}\overline{\square}\Xi_{j)}-\frac{1}{2}\mathring{A}^{\alpha}\mathring{A}_{\alpha}\partial_{t}\partial_{(i}\Xi_{j)} \nonumber\\
        && -\mathring{A}^{k}\mathring{A}^{t}\partial_{i}\partial_{j}\Xi_{k} +\mathring{A}^{k}\partial_{k}\partial_{t}\Xi_{(i}A_{j)}+\mathring{A}^t\mathring{A}_{(i}\overline{\nabla}^2\Xi_{j)}.
\end{eqnarray}
\section{The full forms of the velocities}\label{Appendix2}
In Sec.~\ref{velocitySolutions}, we have obtained all possible velocities of GWs in general Einstein-vector theory. Their specific forms are as follows:
\begin{eqnarray}
    v_1&=&\frac{-2\gamma \mathring{A}^t\mathring{A}^z \pm \sqrt{\big(C_1+2\gamma \mathring{A}^t\mathring{A}^t\big)\big(C_1+2\gamma \mathring{A}^z\mathring{A}^z\big)}}{C_1}, \\
    v_2&=&\frac{-2\gamma^2 \mathring{A}^t\mathring{A}^z \pm \sqrt{\big(C_2+2\gamma^2 \mathring{A}^t\mathring{A}^t\big)\big(C_2+2\gamma^2 \mathring{A}^z\mathring{A}^z\big)}}{C_2},\\
    v_3&=&\frac{-2\gamma(4\gamma-1) \mathring{A}^t\mathring{A}^z \pm \sqrt{\big[C_3+2\gamma(4\gamma-1) \mathring{A}^t\mathring{A}^t\big]\big[C_3+2\gamma(4\gamma-1) \mathring{A}^z\mathring{A}^z\big]}}{C_3},\\
    v_4&=&\frac{\mathring{A}^z}{\mathring{A}^t}.
\end{eqnarray}
Here, 
\begin{eqnarray}
    C_1&=&2 -(2\beta+\gamma)\mathring{A}^{t}\mathring{A}^{t} +(2\beta-\gamma)\big(\mathring{A}^{x}\mathring{A}^{x} +\mathring{A}^{z}\mathring{A}^{z}\big),\\
    C_2&=&2 -(2\beta+\gamma)\mathring{A}^{t}\mathring{A}^{t} +(2\beta+ \gamma-\gamma^2)\big(\mathring{A}^{x}\mathring{A}^{x} +\mathring{A}^{z}\mathring{A}^{z}\big),\\
    C_3&=&6 -3(2\beta+\gamma)\mathring{A}^{t}\mathring{A}^{t} +(6\beta+5\gamma-8\gamma^2)\big(\mathring{A}^{x}\mathring{A}^{x} +\mathring{A}^{z}\mathring{A}^{z}\big).
\end{eqnarray} 
\section{The specific forms of some quantities} \label{Appendix3}
Here, we will represent the specific forms of some quantities that have been used in discussions of polarization modes of GWs.
\begin{eqnarray}
    D_1(\mathring{A}^{t},\mathring{A}^{x},\mathring{A}^{z})&=& 2\gamma\mathring{A}^{z}\mathring{A}^{z} -(2\beta-\gamma) \big(\mathring{A}^{t}\mathring{A}^{t} -\mathring{A}^{x}\mathring{A}^{x} -\mathring{A}^{z}\mathring{A}^{z}\big),\\
    D_2(\mathring{A}^{t},\mathring{A}^{x},\mathring{A}^{z})&=& 2\gamma \mathring{A}^{t}\mathring{A}^{t}+(2\beta-\gamma)\big(\mathring{A}^{t}\mathring{A}^{t} -\mathring{A}^{x}\mathring{A}^{x} -\mathring{A}^z\mathring{A}^{z}\big),\\
    D_3(\mathring{A}^{t},\mathring{A}^{x},\mathring{A}^{z})&=& 2\gamma^2\mathring{A}^{z}\mathring{A}^{z} -(2\beta+\gamma-2\gamma^2) \big(\mathring{A}^{t}\mathring{A}^{t} -\mathring{A}^{x}\mathring{A}^{x} -\mathring{A}^{z}\mathring{A}^{z}\big),\\
    D_4(\mathring{A}^{t},\mathring{A}^{x},\mathring{A}^{z})&=& 2\gamma^2 \mathring{A}^{t}\mathring{A}^{t}+(2\beta+\gamma-2\gamma^2)\big(\mathring{A}^{t}\mathring{A}^{t} -\mathring{A}^{x}\mathring{A}^{x} -\mathring{A}^z\mathring{A}^{z}\big),\\
    D_5(\mathring{A}^{t},\mathring{A}^{x},\mathring{A}^{z})&=&2\gamma (4\gamma-1)\mathring{A}^z\mathring{A}^z-(6\beta+5\gamma-8\gamma^2)\big(\mathring{A}^t\mathring{A}^t-\mathring{A}^x\mathring{A}^x-\mathring{A}^z\mathring{A}^z\big),\\
    D_6(\mathring{A}^{t},\mathring{A}^{x},\mathring{A}^{z})&=&2\gamma (4\gamma-1)\mathring{A}^t\mathring{A}^t+(6\beta+5\gamma-8\gamma^2)\big(\mathring{A}^t\mathring{A}^t-\mathring{A}^x\mathring{A}^x-\mathring{A}^z\mathring{A}^z\big).
\end{eqnarray}
\begin{eqnarray}
    D_1(\mathring{A}^{t},\mathring{A}^{z})&=& 2\gamma\mathring{A}^{z}\mathring{A}^{z} -(2\beta-\gamma) \big(\mathring{A}^{t}\mathring{A}^{t} -\mathring{A}^{z}\mathring{A}^{z}\big),\\
    D_2(\mathring{A}^{t},\mathring{A}^{z})&=& 2\gamma \mathring{A}^{t}\mathring{A}^{t}+(2\beta-\gamma)\big(\mathring{A}^{t}\mathring{A}^{t} -\mathring{A}^z\mathring{A}^{z}\big),\\
    D_3(\mathring{A}^{t},\mathring{A}^{z})&=& 2\gamma^2\mathring{A}^{z}\mathring{A}^{z} -(2\beta+\gamma-2\gamma^2) \big(\mathring{A}^{t}\mathring{A}^{t} -\mathring{A}^{z}\mathring{A}^{z}\big),\\
    D_4(\mathring{A}^{t},\mathring{A}^{z})&=& 2\gamma^2 \mathring{A}^{t}\mathring{A}^{t}+(2\beta+\gamma-2\gamma^2)\big(\mathring{A}^{t}\mathring{A}^{t} -\mathring{A}^z\mathring{A}^{z}\big),\\
    D_5(\mathring{A}^{t},\mathring{A}^{z})&=&2\gamma (4\gamma-1)\mathring{A}^z\mathring{A}^z-(6\beta+5\gamma-8\gamma^2)\big(\mathring{A}^t\mathring{A}^t-\mathring{A}^z\mathring{A}^z\big),\\
    D_6(\mathring{A}^{t},\mathring{A}^{z})&=&2\gamma (4\gamma-1)\mathring{A}^t\mathring{A}^t+(6\beta+5\gamma-8\gamma^2)\big(\mathring{A}^t\mathring{A}^t-\mathring{A}^z\mathring{A}^z\big).
\end{eqnarray}
\begin{eqnarray}
    f^y&=&(\gamma-1)\mathring{A}^x(\partial_z\partial_z -\partial_t\partial_t)h_{\times} +\big[(\gamma-1)\mathring{A}^t\partial_z\partial_z+\mathring{A}^t\partial_t\partial_t+\gamma\mathring{A}^z\partial_t\partial_z\big]\Xi_y+(\partial_{t}\partial_{t} -\partial_{z}\partial_{z})\Sigma_{y}, 
    \\
    F_{ty}&=&-\frac{1}{4}\Big[\big(2+D_1(\mathring{A}^t,\mathring{A}^x,\mathring{A}^z)\big)\partial_{z}\partial_{z}+2\gamma \mathring{A}^{t}\mathring{A}^{z}\partial_{t}\partial_{z}
    \Big]\Xi_y-\frac{1}{2}\gamma \big(\mathring{A}^{t}\partial_{z}\partial_{z} +\mathring{A}^{z}\partial_{t}\partial_{z}\big)\Sigma_{y}, \label{equations-Fg2} 
    \\
    F_{xy}&=&-\frac{1}{4}\Big[\big(2+D_1(\mathring{A}^t,\mathring{A}^x,\mathring{A}^z)\big)\partial_{z}\partial_{z}+4\gamma\mathring{A}^{t}\mathring{A}^{z}\partial_{t}\partial_{z}-\big(2 -D_2(\mathring{A}^t,\mathring{A}^x,\mathring{A}^z)\big)\partial_{t}\partial_{t}\Big]h_{\times} \nonumber\\
    && -\frac{1}{2}\gamma\mathring{A}^{x}\big(\mathring{A}^{t}\partial_{t}\partial_{t}+\mathring{A}^{z}\partial_{t}\partial_z\big)\Xi_{y} +\frac{1}{2}\gamma \mathring{A}^{x}(\partial_{z}\partial_{z} -\partial_{t}\partial_{t})\Sigma_{y},\label{equations-Fg3}
    \\
    F_{yz}&=&-\frac{1}{4}\Big[\big(2+D_1(\mathring{A}^t,\mathring{A}^x,\mathring{A}^z)\big)\partial_{t}\partial_{z}+2\gamma \mathring{A}^{t}\mathring{A}^{z}\partial_{t}\partial_{t}\Big]\Xi_y-\frac{1}{2}\gamma \big(\mathring{A}^{t}\partial_{t}\partial_{z} +\mathring{A}^{t}\partial_{t}\partial_{t}\big)\Sigma_{y}.
\end{eqnarray}

\bibliographystyle{unsrt}
\bibliography{referenceData}

\begin{thebibliography}{100}

\bibitem{LIGOScientific:2016aoc}
B.~P. Abbott et~al.
\newblock {Observation of Gravitational Waves from a Binary Black Hole Merger}.
\newblock {\em Phys. Rev. Lett.}, 116(6):061102, 2016.

\bibitem{LIGOScientific:2016sjg}
B.~P. Abbott et~al.
\newblock {GW151226: Observation of Gravitational Waves from a 22-Solar-Mass
  Binary Black Hole Coalescence}.
\newblock {\em Phys. Rev. Lett.}, 116(24):241103, 2016.

\bibitem{Zhao:2021zlr}
Z.~C. Zhao and Z.~Cao.
\newblock {Stochastic gravitational wave background due to gravitational wave
  memory}.
\newblock {\em Sci. China Phys. Mech. Astron.}, 65(11):119511, 2022.

\bibitem{Bi:2023tib}
Y.~C. Bi, Y.~M. Wu, Z.~C. Chen, and Q.~G. Huang.
\newblock {Implications for the supermassive black hole binaries from the
  NANOGrav 15-year data set}.
\newblock {\em Sci. China Phys. Mech. Astron.}, 66(12):120402, 2023.

\bibitem{Wang:2023len}
Z.~Wang, L.~Lei, H.~Jiao, L.~Feng, and Y.~Z. Fan.
\newblock {The nanohertz stochastic gravitational wave background from cosmic
  string loops and the abundant high redshift massive galaxies}.
\newblock {\em Sci. China Phys. Mech. Astron.}, 66(12):120403, 2023.

\bibitem{Wang:2021srv}
L.~F. Wang, S.~J. Jin, J.~F. Zhang, and X.~Zhang.
\newblock {Forecast for cosmological parameter estimation with
  gravitational-wave standard sirens from the LISA-Taiji network}.
\newblock {\em Sci. China Phys. Mech. Astron.}, 65(1):210411, 2022.

\bibitem{Gao:2022hho}
Q.~Gao.
\newblock {Constraint on the mass of graviton with gravitational waves}.
\newblock {\em Sci. China Phys. Mech. Astron.}, 66(2):220411, 2023.

\bibitem{Shan:2022xfx}
X.~Shan, G.~Li, X.~Chen, W.~Zheng, and W.~Zhao.
\newblock {Wave effect of gravitational waves intersected with a microlens
  field: A new algorithm and supplementary study}.
\newblock {\em Sci. China Phys. Mech. Astron.}, 66(3):239511, 2023.

\bibitem{LIGOScientific:2017bnn}
B.~P. Abbott et~al.
\newblock {GW170104: Observation of a 50-Solar-Mass Binary Black Hole
  Coalescence at Redshift 0.2}.
\newblock {\em Phys. Rev. Lett.}, 118(22):221101, 2017.
\newblock [Erratum: Phys.Rev.Lett. 121, 129901 (2018)].

\bibitem{LIGOScientific:2017ycc}
B.~P. Abbott et~al.
\newblock {GW170814: A Three-Detector Observation of Gravitational Waves from a
  Binary Black Hole Coalescence}.
\newblock {\em Phys. Rev. Lett.}, 119(14):141101, 2017.

\bibitem{LIGOScientific:2017vwq}
B.~P. Abbott et~al.
\newblock {GW170817: Observation of Gravitational Waves from a Binary Neutron
  Star Inspiral}.
\newblock {\em Phys. Rev. Lett.}, 119(16):161101, 2017.

\bibitem{LIGOScientific:2021qlt}
R.~Abbott et~al.
\newblock {Observation of Gravitational Waves from Two Neutron
  Star\textendash{}Black Hole Coalescences}.
\newblock {\em Astrophys. J. Lett.}, 915(1):L5, 2021.

\bibitem{LIGOScientific:2018mvr}
B.~P. Abbott et~al.
\newblock {GWTC-1: A Gravitational-wave Transient Catalog of Compact Binary
  Mergers Observed by LIGO and Virgo during the First and Second Observing
  Runs}.
\newblock {\em Phys. Rev. X}, 9(3):031040, 2019.

\bibitem{LIGOScientific:2020ibl}
R.~Abbott et~al.
\newblock {GWTC-2: Compact Binary Coalescences Observed by LIGO and Virgo
  during the First Half of the Third Observing Run}.
\newblock {\em Phys. Rev. X}, 11:021053, 2021.

\bibitem{LIGOScientific:2021djp}
R.~Abbott et~al.
\newblock {GWTC-3: Compact Binary Coalescences Observed by LIGO and Virgo
  During the Second Part of the Third Observing Run}.
\newblock {\em arXiv:2111.03606}, 2021.

\bibitem{NANOGrav:2023gor}
G.~Agazie et~al.
\newblock {The NANOGrav 15 yr Data Set: Evidence for a Gravitational-wave
  Background}.
\newblock {\em Astrophys. J. Lett.}, 951(1):L8, 2023.

\bibitem{EPTA:2023fyk}
J.~Antoniadis et~al.
\newblock {The second data release from the European Pulsar Timing Array III.
  Search for gravitational wave signals}.
\newblock {\em Astron. Astrophys.}, 678:A50, 2023.

\bibitem{Reardon:2023gzh}
D.~J. Reardon et~al.
\newblock {Search for an Isotropic Gravitational-wave Background with the
  Parkes Pulsar Timing Array}.
\newblock {\em Astrophys. J. Lett.}, 951(1):L6, 2023.

\bibitem{Xu:2023wog}
H.~Xu et~al.
\newblock {Searching for the Nano-Hertz Stochastic Gravitational Wave
  Background with the Chinese Pulsar Timing Array Data Release I}.
\newblock {\em Res. Astron. Astrophys.}, 23(7):075024, 2023.

\bibitem{Smith:1936mlg}
S.~Smith.
\newblock {The Mass of the Virgo Cluster}.
\newblock {\em Astrophys. J.}, 83:23--30, 1936.

\bibitem{zwicky1933helvetica}
F~Zwicky.
\newblock Helvetica physica acta 6.
\newblock {\em 110{\^a} A S}, 127, 1933.

\bibitem{Zwicky:1937zza}
F.~Zwicky.
\newblock {On the Masses of Nebulae and of Clusters of Nebulae}.
\newblock {\em Astrophys. J.}, 86:217--246, 1937.

\bibitem{Peebles:2002gy}
P.~J.~E. Peebles and B.~Ratra.
\newblock {The Cosmological Constant and Dark Energy}.
\newblock {\em Rev. Mod. Phys.}, 75:559--606, 2003.

\bibitem{Goroff:1985th}
M.~H. Goroff and A.~Sagnotti.
\newblock {The Ultraviolet Behavior of Einstein Gravity}.
\newblock {\em Nucl. Phys. B}, 266:709--736, 1986.

\bibitem{tHooft:1974toh}
G.~'t~Hooft and M.~J.~G. Veltman.
\newblock {One loop divergencies in the theory of gravitation}.
\newblock {\em Ann. Inst. H. Poincare Phys. Theor. A}, 20:69--94, 1974.

\bibitem{SupernovaCosmologyProject:1998vns}
S.~Perlmutter et~al.
\newblock {Measurements of $\Omega$ and $\Lambda$ from 42 high redshift
  supernovae}.
\newblock {\em Astrophys. J.}, 517:565--586, 1999.

\bibitem{La:1989za}
D.~La and P.~J. Steinhardt.
\newblock {Extended Inflationary Cosmology}.
\newblock {\em Phys. Rev. Lett.}, 62:376, 1989.
\newblock [Erratum: Phys.Rev.Lett. 62, 1066 (1989)].

\bibitem{PhysRev.124.925}
C.~Brans and R.~H. Dicke.
\newblock Mach's principle and a relativistic theory of gravitation.
\newblock {\em Phys. Rev.}, 124:925--935, Nov 1961.

\bibitem{Horndeski:1974wa}
G.~W. Horndeski.
\newblock {Second-order scalar-tensor field equations in a four-dimensional
  space}.
\newblock {\em Int. J. Theor. Phys.}, 10:363--384, 1974.

\bibitem{Fu:2019xtx}
Q.~M. Fu, H.~Yu, L.~Zhao, and Y.~X. Liu.
\newblock {Thick brane in reduced Horndeski theory}.
\newblock {\em Phys. Rev. D}, 100(12):124057, 2019.

\bibitem{DeFelice:2010jn}
A.~De~Felice and S.~Tsujikawa.
\newblock {Generalized Brans-Dicke theories}.
\newblock {\em JCAP}, 07:024, 2010.

\bibitem{Jacobson:2007veq}
T.~Jacobson.
\newblock {Einstein-aether gravity: A status report}.
\newblock {\em PoS}, QG-PH:020, 2007.

\bibitem{Tasinato:2014eka}
G.~Tasinato.
\newblock {Cosmic Acceleration from Abelian Symmetry Breaking}.
\newblock {\em JHEP}, 04:067, 2014.

\bibitem{Heisenberg:2014rta}
L.~Heisenberg.
\newblock {Generalization of the Proca Action}.
\newblock {\em JCAP}, 05:015, 2014.

\bibitem{Kostelecky:2003fs}
V.~A. Kostelecky.
\newblock {Gravity, Lorentz violation, and the standard model}.
\newblock {\em Phys. Rev. D}, 69:105009, 2004.

\bibitem{Geng:2015kvs}
W.~J. Geng and H.~Lu.
\newblock {Einstein-Vector Gravity, Emerging Gauge Symmetry and de Sitter
  Bounce}.
\newblock {\em Phys. Rev. D}, 93(4):044035, 2016.

\bibitem{Bekenstein:2004ne}
J.~D. Bekenstein.
\newblock {Relativistic gravitation theory for the MOND paradigm}.
\newblock {\em Phys. Rev. D}, 70:083509, 2004.
\newblock [Erratum: Phys.Rev.D 71, 069901 (2005)].

\bibitem{Moffat:2005si}
J.~W. Moffat.
\newblock {Scalar-tensor-vector gravity theory}.
\newblock {\em JCAP}, 03:004, 2006.

\bibitem{Bauer:2008zj}
F.~Bauer and D.~A. Demir.
\newblock {Inflation with Non-Minimal Coupling: Metric versus Palatini
  Formulations}.
\newblock {\em Phys. Lett. B}, 665:222--226, 2008.

\bibitem{Markkanen:2017tun}
T.~Markkanen, T.~Tenkanen, V.~Vaskonen, and H.~Veerm\"ae.
\newblock {Quantum corrections to quartic inflation with a non-minimal
  coupling: metric vs. Palatini}.
\newblock {\em JCAP}, 03:029, 2018.

\bibitem{Vollick:2003aw}
D.~N. Vollick.
\newblock {1/R Curvature corrections as the source of the cosmological
  acceleration}.
\newblock {\em Phys. Rev. D}, 68:063510, 2003.

\bibitem{Bauer:2010jg}
F.~Bauer and D.~A. Demir.
\newblock {Higgs-Palatini Inflation and Unitarity}.
\newblock {\em Phys. Lett. B}, 698:425--429, 2011.

\bibitem{Rubio:2019ypq}
J.~Rubio and E.~S. Tomberg.
\newblock {Preheating in Palatini Higgs inflation}.
\newblock {\em JCAP}, 04:021, 2019.

\bibitem{Olmo:2011uz}
G.~J. Olmo.
\newblock {Palatini Approach to Modified Gravity: f(R) Theories and Beyond}.
\newblock {\em Int. J. Mod. Phys. D}, 20:413--462, 2011.

\bibitem{BeltranJimenez:2018vdo}
J.~Beltr\'an~Jim\'enez, L.~Heisenberg, and T.~S. Koivisto.
\newblock {Teleparallel Palatini theories}.
\newblock {\em JCAP}, 08:039, 2018.

\bibitem{Dong:2021jtd}
Y.~Q. Dong and Y.~X. Liu.
\newblock {Polarization modes of gravitational waves in Palatini-Horndeski
  theory}.
\newblock {\em Phys. Rev. D}, 105(6):064035, 2022.

\bibitem{Lu:2020eux}
J.~Lu, J.~Li, H.~Guo, Z.~Zhuang, and X.~Zhao.
\newblock {Linearized physics and gravitational-waves polarizations in the
  Palatini formalism of GBD theory}.
\newblock {\em Phys. Lett. B}, 811:135985, 2020.

\bibitem{Kubota:2020ehu}
M.~Kubota, K.~Y. Oda, K.~Shimada, and M.~Yamaguchi.
\newblock {Cosmological Perturbations in Palatini Formalism}.
\newblock {\em JCAP}, 03:006, 2021.

\bibitem{Sotiriou:2008rp}
T.~P. Sotiriou and V.~Faraoni.
\newblock {f(R) Theories Of Gravity}.
\newblock {\em Rev. Mod. Phys.}, 82:451--497, 2010.

\bibitem{Cui:2020fiz}
Z.~Q. Cui, Z.~C. Lin, J.~J. Wan, Y.~X. Liu, and L.~Zhao.
\newblock {Tensor Perturbations and Thick Branes in Higher-dimensional $f(R)$
  Gravity}.
\newblock {\em JHEP}, 12:130, 2020.

\bibitem{Baibosunov:1990qm}
M.~B. Baibosunov, V.~Ts. Gurovich, and U.~M. Imanaliev.
\newblock {Model of the early universe in f(R) theory}.
\newblock {\em Sov. Phys. JETP}, 71:636--642, 1990.

\bibitem{Chen:2020zzs}
J.~Chen, W.~D. Guo, and Y.~X. Liu.
\newblock {Thick branes with inner structure in mimetic f(R) gravity}.
\newblock {\em Eur. Phys. J. C}, 81(8):709, 2021.

\bibitem{Kaluza:1921tu}
T.~Kaluza.
\newblock Zum unit\"atsproblem der physik.
\newblock {\em Sitzungsber. Preuss. Akad. Wiss. Berlin (Math. Phys. )},
  1921:966--972, 1921.

\bibitem{Rubakov:1983bb}
V.~A. Rubakov and M.~E. Shaposhnikov.
\newblock {Do We Live Inside a Domain Wall?}
\newblock {\em Phys. Lett. B}, 125:136--138, 1983.

\bibitem{Tan:2020sys}
Q.~Tan, W.~D. Guo, Y.~P. Zhang, and Y.~X. Liu.
\newblock {Gravitational resonances on $f(T)$-branes}.
\newblock {\em Eur. Phys. J. C}, 81(4):373, 2021.

\bibitem{Arkani-Hamed:1998jmv}
N.~Arkani-Hamed, S.~Dimopoulos, and G.~R. Dvali.
\newblock {The Hierarchy problem and new dimensions at a millimeter}.
\newblock {\em Phys. Lett. B}, 429:263--272, 1998.

\bibitem{Randall:1999vf}
L.~Randall and R.~Sundrum.
\newblock {An alternative to compactification}.
\newblock {\em Phys. Rev. Lett.}, 83:4690--4693, 1999.

\bibitem{Yu:2019jlb}
H.~Yu, Z.~C. Lin, and Y.~X. Liu.
\newblock {Gravitational waves and extra dimensions: a short review}.
\newblock {\em Commun. Theor. Phys.}, 71(8):991--1006, 2019.

\bibitem{Li:2022kly}
C.~C. Li, Z.~Q. Cui, T.~T. Sui, and Y.~X. Liu.
\newblock {Effective action of a self-interacting scalar field on brane}.
\newblock {\em Eur. Phys. J. C}, 83(2):119, 2023.

\bibitem{Lin:2022hus}
Z.~C. Lin, H.~Yu, and Y.~X. Liu.
\newblock {Shortcut in codimension-2 brane cosmology in light of GW170817}.
\newblock {\em Eur. Phys. J. C}, 83(3):190, 2023.

\bibitem{Xu:2022xxd}
N.~Xu, J.~Chen, Y.~P. Zhang, and Y.~X. Liu.
\newblock {Multikink brane in Gauss-Bonnet gravity and its stability}.
\newblock {\em Phys. Rev. D}, 107(12):124011, 2023.

\bibitem{Zhong:2022wlw}
Y.~Zhong, K.~Yang, and Y.~X. Liu.
\newblock {Thick brane in Rastall gravity}.
\newblock {\em JHEP}, 09:128, 2022.

\bibitem{Goldstein:2017mmi}
A.~Goldstein et~al.
\newblock {An Ordinary Short Gamma-Ray Burst with Extraordinary Implications:
  Fermi-GBM Detection of GRB 170817A}.
\newblock {\em Astrophys. J. Lett.}, 848(2):L14, 2017.

\bibitem{Savchenko:2017ffs}
V.~Savchenko et~al.
\newblock {INTEGRAL Detection of the First Prompt Gamma-Ray Signal Coincident
  with the Gravitational-wave Event GW170817}.
\newblock {\em Astrophys. J. Lett.}, 848(2):L15, 2017.

\bibitem{Eardley:1973zuo}
D.~M. Eardley, D.~L. Lee, and A.~P. Lightman.
\newblock {Gravitational-wave observations as a tool for testing relativistic
  gravity}.
\newblock {\em Phys. Rev. D}, 8:3308--3321, 1973.

\bibitem{Eardley:1973br}
D.~M. Eardley, D.~L. Lee, A.~P. Lightman, R.~V. Wagoner, and C.~M. Will.
\newblock {Gravitational-wave observations as a tool for testing relativistic
  gravity}.
\newblock {\em Phys. Rev. Lett.}, 30:884--886, 1973.

\bibitem{Hyun:2018pgn}
Y.~H. Hyun, Y.~Kim, and S.~Lee.
\newblock {Exact amplitudes of six polarization modes for gravitational waves}.
\newblock {\em Phys. Rev. D}, 99(12):124002, 2019.

\bibitem{Bardeen:1980kt}
J.~M. Bardeen.
\newblock {Gauge Invariant Cosmological Perturbations}.
\newblock {\em Phys. Rev. D}, 22:1882--1905, 1980.

\bibitem{Flanagan:2005yc}
E.~E. Flanagan and S.~A. Hughes.
\newblock {The basics of gravitational wave theory}.
\newblock {\em New J. Phys.}, 7:204, 2005.

\bibitem{Liang:2022hxd}
D.~Liang, R.~Xu, X.~Lu, and L.~Shao.
\newblock {Polarizations of gravitational waves in the bumblebee gravity
  model}.
\newblock {\em Phys. Rev. D}, 106(12):124019, 2022.

\bibitem{Dong:2023xyb}
Y.~Q. Dong, Y.~Q. Liu, and Y.~X. Liu.
\newblock {Polarization modes of gravitational waves in generalized Proca
  theory}.
\newblock {\em Phys. Rev. D}, 109(2):024014, 2024.

\bibitem{Liang:2017ahj}
D.~Liang, Y.~Gong, S.~Hou, and Y.~Liu.
\newblock {Polarizations of gravitational waves in $f(R)$ gravity}.
\newblock {\em Phys. Rev. D}, 95(10):104034, 2017.

\bibitem{Gong:2018vbo}
Y.~Gong, S.~Hou, E.~Papantonopoulos, and D.~Tzortzis.
\newblock {Gravitational waves and the polarizations in Ho\v{r}ava gravity
  after GW170817}.
\newblock {\em Phys. Rev. D}, 98(10):104017, 2018.

\bibitem{Wagle:2019mdq}
P.~Wagle, A.~Saffer, and N.~Yunes.
\newblock {Polarization modes of gravitational waves in Quadratic gravity}.
\newblock {\em Phys. Rev. D}, 100(12):124007, 2019.

\bibitem{Bahamonde:2021dqn}
S.~Bahamonde, M.~Caruana, K.~F. Dialektopoulos, V.~Gakis, M.~Hohmann,
  J.~Levi~Said, E.~N. Saridakis, and J.~Sultana.
\newblock {Gravitational-wave propagation and polarizations in the teleparallel
  analog of Horndeski gravity}.
\newblock {\em Phys. Rev. D}, 104(8):084082, 2021.

\bibitem{Dong:2023bgt}
Y.~Q. Dong, Y.~Q. Liu, and Y.~X. Liu.
\newblock {Polarization modes of gravitational waves in general modified
  gravity: General metric theory and general scalar-tensor theory}.
\newblock {\em Phys. Rev. D}, 109(4):044013, 2024.

\bibitem{Liu:2022cwb}
Y.~Q. Liu, Y.~Q. Dong, and Y.~X. Liu.
\newblock {Classification of gravitational waves in higher-dimensional
  space-time and possibility of observation}.
\newblock {\em Eur. Phys. J. C}, 83(9):857, 2023.

\bibitem{Dong:2022cvf}
Y.~Q. Dong, Y.~Q. Liu, and Y.~X. Liu.
\newblock {Constraining Palatini\textendash{}Horndeski theory with
  gravitational waves after GW170817}.
\newblock {\em Eur. Phys. J. C}, 83(8):702, 2023.

\bibitem{Kuroda:2015owv}
K.~Kuroda, W.~T. Ni, and W.~P. Pan.
\newblock {Gravitational waves: Classification, Methods of detection,
  Sensitivities, and Sources}.
\newblock {\em Int. J. Mod. Phys. D}, 24(14):1530031, 2015.

\bibitem{Isi:2017equ}
M.~Isi, M.~Pitkin, and A.~J. Weinstein.
\newblock {Probing Dynamical Gravity with the Polarization of Continuous
  Gravitational Waves}.
\newblock {\em Phys. Rev. D}, 96(4):042001, 2017.

\bibitem{Niu:2018oox}
R.~Niu and W.~Zhao.
\newblock {Constraining the non-Einsteinian polarizations of gravitational
  waves by pulsar timing array}.
\newblock {\em Sci. China Phys. Mech. Astron.}, 62(7):970411, 2019.

\bibitem{Hou:2024xbv}
S.~Hou, X.~L. Fan, T.~Zhu, and Z.~H. Zhu.
\newblock {Nontensorial gravitational wave polarizations from the tensorial
  degrees of freedom: Linearized Lorentz-violating theory of gravity}.
\newblock {\em Phys. Rev. D}, 109(8):084011, 2024.

\bibitem{Battista:2021rlh}
E.~Battista and V.~De~Falco.
\newblock {First post-Newtonian generation of gravitational waves in
  Einstein-Cartan theory}.
\newblock {\em Phys. Rev. D}, 104(8):084067, 2021.

\bibitem{Gao:2021vxb}
Q.~Gao.
\newblock {Primordial black holes and secondary gravitational waves from
  chaotic inflation}.
\newblock {\em Sci. China Phys. Mech. Astron.}, 64(8):280411, 2021.

\bibitem{Wu:2023hsa}
Y.~M. Wu, Z.~C. Chen, and Q.~G. Huang.
\newblock {Cosmological interpretation for the stochastic signal in pulsar
  timing arrays}.
\newblock {\em Sci. China Phys. Mech. Astron.}, 67(4):240412, 2024.

\bibitem{Yi:2023mbm}
Z.~Yi, Q.~Gao, Y.~Gong, Y.~Wang, and F.~Zhang.
\newblock {Scalar induced gravitational waves in light of Pulsar Timing Array
  data}.
\newblock {\em Sci. China Phys. Mech. Astron.}, 66(12):120404, 2023.

\bibitem{Bian:2021ini}
L.~Bian et~al.
\newblock {The Gravitational-wave physics II: Progress}.
\newblock {\em Sci. China Phys. Mech. Astron.}, 64(12):120401, 2021.

\bibitem{Schumacher:2023cxh}
K.~Schumacher, S.~E. Perkins, A.~Shaw, K.~Yagi, and N.~Yunes.
\newblock {Gravitational wave constraints on Einstein-\ae{}ther theory with
  LIGO/Virgo data}.
\newblock {\em Phys. Rev. D}, 108(10):104053, 2023.

\bibitem{Chen:2023uiz}
Z.~C. Chen, Y.~M. Wu, Y.~C. Bi, and Q.~G. Huang.
\newblock {Search for nontensorial gravitational-wave backgrounds in the
  NANOGrav 15-year dataset}.
\newblock {\em Phys. Rev. D}, 109(8):084045, 2024.

\bibitem{lovelock1971einstein}
D.~Lovelock.
\newblock {The Einstein tensor and its generalizations}.
\newblock {\em J. Math. Phys.}, 12:498--501, 1971.

\bibitem{jackiw2003chern}
R.~Jackiw and S.~Y. Pi.
\newblock {Chern-Simons modification of general relativity}.
\newblock {\em Phys. Rev. D}, 68:104012, 2003.

\bibitem{bluhm2008spontaneous}
R.~Bluhm, S.~H. Fung, and V.~A. Kostelecky.
\newblock {Spontaneous Lorentz and Diffeomorphism Violation, Massive Modes, and
  Gravity}.
\newblock {\em Phys. Rev. D}, 77:065020, 2008.

\bibitem{Gong:2018cgj}
Y.~Gong, S.~Hou, D.~Liang, and E.~Papantonopoulos.
\newblock {Gravitational waves in Einstein-\ae{}ther and generalized TeVeS
  theory after GW170817}.
\newblock {\em Phys. Rev. D}, 97(8):084040, 2018.

\bibitem{Lin:2018ken}
K.~Lin, X.~Zhao, C.~Zhang, T.~Liu, B.~Wang, S.~Zhang, X.~Zhang, W.~Zhao,
  T.~Zhu, and A.~Wang.
\newblock {Gravitational waveforms, polarizations, response functions, and
  energy losses of triple systems in Einstein-aether theory}.
\newblock {\em Phys. Rev. D}, 99(2):023010, 2019.

\bibitem{Jacobson:2004ts}
T.~Jacobson and D.~Mattingly.
\newblock {Einstein-\ae{}ther waves}.
\newblock {\em Phys. Rev. D}, 70:024003, 2004.

\bibitem{abbott2017gravitational}
B.~P. Abbott et~al.
\newblock {Gravitational Waves and Gamma-rays from a Binary Neutron Star
  Merger: GW170817 and GRB 170817A}.
\newblock {\em Astrophys. J. Lett.}, 848(2):L13, 2017.

\bibitem{abbott2019tests}
B.~P. Abbott et~al.
\newblock {Tests of General Relativity with GW170817}.
\newblock {\em Phys. Rev. Lett.}, 123(1):011102, 2019.

\bibitem{NANOGrav:2023ygs}
G.~Agazie et~al.
\newblock {The NANOGrav 15 yr Data Set: Search for Transverse Polarization
  Modes in the Gravitational-wave Background}.
\newblock {\em Astrophys. J. Lett.}, 964(1):L14, 2024.

\bibitem{Papallo:2017qvl}
G.~Papallo and H.~S. Reall.
\newblock {On the local well-posedness of Lovelock and Horndeski theories}.
\newblock {\em Phys. Rev. D}, 96(4):044019, 2017.

\bibitem{kreiss1989initial}
H.~O. Kreiss and J.~Lorenz.
\newblock {\em Initial-boundary Value Problems and the Navier-Stokes
  Equations}.
\newblock Number v. 136 in Initial-boundary value problems and the
  Navier-Stokes equations. Academic Press, 1989.

\bibitem{Sarbach:2012pr}
O.~Sarbach and M.~Tiglio.
\newblock {Continuum and Discrete Initial-Boundary-Value Problems and
  Einstein's Field Equations}.
\newblock {\em Living Rev. Rel.}, 15:9, 2012.

\bibitem{Davies:2021frz}
I.~Davies and H.~S. Reall.
\newblock {Well-posed formulation of Einstein-Maxwell effective field theory}.
\newblock {\em Phys. Rev. D}, 106(10):104019, 2022.

\bibitem{LISA:2017pwj}
P.~Amaro-Seoane et~al.
\newblock {Laser Interferometer Space Antenna}.
\newblock {\em arXiv:1702.00786}, 2017.

\bibitem{10.1093.ptep.ptaa083}
Z.~Luo, Y.~Wang, Y.~Wu, W.~Hu, and G.~Jin.
\newblock {The Taiji program: A concise overview}.
\newblock {\em PTEP}, 2021(5):05A108, 2021.

\bibitem{TianQin:2015yph}
J.~Luo et~al.
\newblock {TianQin: a space-borne gravitational wave detector}.
\newblock {\em Class. Quant. Grav.}, 33(3):035010, 2016.

\end{thebibliography}

\end{document}